%% file: main.tex
\documentclass[referee]{aa}
\def\twocol{"n"}
\usepackage[utf8]{inputenc}
\usepackage[english]{babel}
\usepackage{graphicx}
\usepackage{float}
\usepackage{amsmath}
\usepackage{fullpage}
\usepackage{caption}
\usepackage[format=hang]{subfig}
\usepackage{lipsum}
\usepackage{natbib}
\usepackage[flushleft]{threeparttable}
\usepackage{ifthen}
\usepackage{placeins}
\bibpunct{(}{)}{;}{a}{}{,}
\AddToHook{begindocument/before}{\usepackage{hyperref}}
\usepackage{xr}
\makeatletter
\newcommand*{\addFileDependency}[1]{
\typeout{(#1)}
\@addtofilelist{#1}
\IfFileExists{#1}{}{\typeout{No file #1.}}
}\makeatother
\newcommand*{\myexternaldocument}[1]{%
	\externaldocument{#1}%
	\addFileDependency{#1.tex}%
	\addFileDependency{#1.aux}%
}
\myexternaldocument{supplementary_material}
\setcitestyle{citesep={;}}
\begin{document}
\nolinenumbers
\title{Study of Io's sodium jets with the TRAPPIST telescopes}
\author{
    A. De Becker\inst{1}\fnmsep\inst{2}\fnmsep\thanks{These authors contributed equally to this work and share first authorship.}
    \and L. A. Head\inst{1}\fnmsep\inst{\star}\fnmsep\thanks{Corresponding author: LA.Head@uliege.be (email address)}
    \and B. Bonfond \inst{1}
    \and E. Jehin \inst{1}
    \and J. Manfroid \inst{1}
    \and Z. Yao \inst{3}
    \and B. Zhang \inst{2}
    \and D. Grodent \inst{1}
    \and N. Schneider \inst{4}
    \and Z. Benkhaldoun \inst{5}
}
\institute{
    Space sciences, Technologies and Astrophysics Research (STAR) Institute, University of Li\`{e}ge, Li\`{e}ge, Belgium
    \and Department of Earth Sciences, University of Hong Kong, Hong Kong, China
    \and Key Laboratory of Earth and Planetary Physics, Institute of Geology and Geophysics, Chinese Academy of Sciences, Beijing, China
    \and Laboratory for Atmospheric and Space Physics, University of Colorado, Boulder, USA
    \and Oukaimeden Observatory, High Energy Physics and Astrophysics Laboratory, Cadi Ayyad University, Marrakech, Morocco 
}
\abstract{
    Io is the most volcanically active body in  the Solar System.
    This volcanic activity results in the ejection of material into Io's atmosphere, which may then escape from the atmosphere to form various structures in the jovian magnetosphere, including the plasma torus and clouds of neutral particles.
    The physical processes involved in the escape of particles - for example, how the volcanoes of Io provide material to the plasma torus - are not yet fully understood.
    In particular, it is not clear to what extent the sodium jet, one of the sodium neutral clouds related to Io, is a proxy of processes that populate the various reservoirs of plasma in Jupiter's magnetosphere.
    Here, we report on observations carried out over 17 nights in 2014-2015, 30 nights in 2021, and 23 nights in 2022-2023 with the TRAPPIST telescopes, in which particular attention was paid to the sodium jet and the quantification of their physical properties (length, brightness).
    It was found that these properties can vary greatly from one jet to another and independently of the position of Io in its orbit.
    No clear link was found between the presence of jets and global brightening of the plasma torus and extended sodium nebula, indicating that jets do not contribute straightforwardly to their population.
    This work also demonstrates the advantage of regular and long-term monitoring to understanding the variability of the sodium jet and presents a large corpus of jet detections against which work in related fields may compare.
}
\maketitle
\section{Introduction}\FloatBarrier

Jupiter is the most massive planet in the Solar System and exerts therefore a considerable influence on its satellite system and, in particular, on its four Galilean moons. Because of an orbital resonance between the three inner moons (Io, Europa, and Ganymede), Io's orbit remains elliptic; this orbital eccentricity, combined with the powerful gravitational field of Jupiter, results in strong tidal heating of Io’s interior which gives rise to Io's intense volcanism \citep{dekleerEA2020}.

This volcanism is an important source of material for the tenuous and patchy atmosphere of Io. This atmosphere is mostly made of SO$_2$ and originates partly from direct outgassing from volcanoes and partly from the sublimation of frost from the surface of Io. Studies indicate that sublimation is the main source of Io's atmosphere \citep{lellouch2005}, but the effect of the volcanic activity of Io cannot be neglected and the exact contribution of both phenomena in atmosphere formation is not yet clearly established. Volcanoes on Io take various forms, including lava lakes and  $>300$  km high plumes \citep{geisslerandgoldstein2007,williamsandhowell2007} and possibly stealth volcanoes \citep{depaterEA2020}. Some of this volcanism appears to vary cyclically with Io’s orbital motion \citep{dekleerEA2020}.

Due to Io's comparatively weak gravitational field and intense bombardment from magnetospheric particles, atmospheric particles are prone to escape. While part of these escaping particles are ionised and will form the plasma torus along Io’s orbit, the remaining particles will stay neutral and form neutral clouds \citep{bagenalanddols2020}. Some ions in the plasma torus will interact with Io’s atmosphere (as the plasma torus rotates faster than Io) and cause atmospheric sputtering, which will then also contribute to the neutral clouds via collisions, charge exchanges, dissociation, and recombination \citep[e.g.][]{summersEA1989, schneiderEA1991, smyth1992, dolsEA2008}.

There are several neutral clouds, including the ``banana'', the streams and the jet \citep{wilsonEA2002}, which are formed by different populations of neutral particles. Though sodium is only a trace component \citep{thomasEA2004}, an emission line caused by resonant scattering \citep{bergstralhEA1975} in the visible band (589 nm; the sodium D-doublet) is by far the brightest emission of the elements present and thus sodium is the easiest particle to detect in the neutral clouds. Sodium may then be used as proxy to estimate the behaviour of the other component, though the extent to which this is the case is disputed \citep{schneiderandbagenal2007}. The fast-moving sodium contained within the streams leaving Io will form an extended sodium nebula \citep{mendilloEA1990}, which extends out to several hundreds of jovian radii and forms a corona around Jupiter.

Part of the atmospheric escape can be attributed to spectacular jets observable by their sodium emission. These jets originate in the exosphere of Io as ions are picked up and rapidly re-neutralised, before escaping the neighbourhood of Io. Upon neutralisation, these atoms retain their original plasma-corotation velocities, with an additional component from their gyromotion, and are thus rapidly ejected in the plane perpendicular to the local magnetic field at Io, with speeds close to the magnetospheric corotation speed relative to Io of 57 km/s and in an anti-jovian fan-like shape from the direction of movement of Io \citep{wilsonEA2002}. The exact relation between these jets, the extended sodium cloud, and the mass loading of the plasma torus remains unclear.

Unlike jets, the banana is a structure in the neutral sodium cloud that is composed of particles that remained neutral and escaped slowly from the atmosphere of Io. Since the banana is not directly controlled by the magnetic field of Jupiter, it extends in the orbital plane of Io in a curved arc that precedes Io in its orbit \citep{wilsonEA2002}. 
 
Due to its intense volcanism and particle escape from its atmosphere, Io is the main source of material for the magnetosphere of Jupiter and plays therefore a crucial role in magnetospheric processes. Studies of the atmosphere of Io and studies of the plasma torus may lead to apparent contradictions; the atmosphere of Io appears stable \citep{rothEA2020}, whereas the plasma torus shows variation that can be explained by variability in Io's volcanic activity \citep{yoshiokaEA2018}. Since the atmosphere represents an intermediate stage between generation on the surface and injection into the plasma torus for iogenic material, it would be expected that the atmosphere be affected in a similar manner as the plasma torus by variable volcanic activity. Studying the neutral sodium clouds and the structures therewithin may shed light on the surface-atmosphere-magnetosphere coupling. 

In this study, we report observations of the neutral sodium clouds carried out by the TRAPPIST telescopes, with particular attention paid to the neutral sodium jets. The purpose of these observations is to characterise the variability of the jets by measuring their size and brightness, as well as understanding the variation in their geometry, in order to improve understanding of Jovian magnetospheric dynamics, particularly of the particle sources from the Io plasma torus.

\section{Telescopes and methods}

\subsection{Data acquisition}

The image data used in this work were collected using the two TRAPPIST (TRAnsiting Planets and PlanetesImals Small Telescope) telescopes, one located at the La Silla observatory in Chile (TRAPPIST-South) and one at the Oukaimeden observatory in Morocco (TRAPPIST-North), both 60-cm robotic Ritchey-Cretien telescopes \citep{jehinEA2011}. TRAPPIST-South and TRAPPIST-North have fields of view of 22$'\times$22$’$ and 20$’\times$20$’$ and a pixel scale of 0.64 $"$px$^{-1}$ and 0.60 $"$px$^{-1}$ respectively. They are both equipped with a narrow-band Na-I filter (central wavelength: 589 nm, FWHM: 3 nm) made by Custom Scientific Inc. which allows for the observation of the sodium D-doublet emission in the vicinity of Io and hence any structures in the neutral sodium cloud.

Observations with TRAPPIST-South were made over seventeen nights between 2014-12-04 and 2015-04-01. These observations were intended as test observations and thus there was no specific restriction on the period of observation; as a consequence, in some of these observations, Io was not in an ideal configuration to detect jet-like structures in the neutral sodium cloud. These images have undergone preliminary analysis in previous work \citep{spiegeleire2019} upon which this work has developed using new methods created for later observations of Io. New observations began in April 2021 with TRAPPIST-South and in June 2021 with TRAPPIST-North. For these observations, the periods of observation were chosen to ensure that Io was far from Jupiter in the observation plane and thus minimally affected by the light reflected from the planet. These observations were performed over 30 nights. Typically, a sequence of five image exposures followed by five bias and five dark exposures (to remove artefacts on the CCD chip from the heavy image saturation) were made with exposure times of 5, 15, and 60 seconds. Flat-field exposures with the sodium filter were taken at the end of each night that an observation of Io was made, in a series of seven dithered frames on the bright sky of the nautical twilight.

An overview of the observations from 2014-2015 are given in table \ref{tab:obs_2014}, and observations from the 2021 and 2022-2023 viewing campaigns are listed in tables \ref{tab:obs_2021} and \ref{tab:obs_2022} respectively.

\subsection{Determination of the brightness}
\label{sec:brightness_conversion}

To allow for physical comparison of the brightness of the structures detected in these images, it is necessary to convert the image units from ADU to rayleigh, which can be performed with the aid of standard stars. 
For example, for the 2014-2015 viewing campaign, the standard star HD72526, with a magnitude of 7.92, was measured to have a mean flux per second of 17678.6 ADU s$^{-1}$, while using the same instrumental setting as for our observations of Io. 
It is known that a magnitude-0 star has a spectral flux of 2.75$\times$10$^{-9}$ erg s$^{-1}$ cm$^{-2}$ \AA$^{-1}$, and hence, using the FWHM of the sodium filter (33 \AA), a flux of 9.08$\times$10$^{-8}$ erg s$^{-1}$ cm$^{-2}$. 
HD72526, a magnitude-8 star, has therefore a flux of 5.73$\times$10$^{-11}$ erg s$^{-1}$ cm$^{-2}$.
Hence, a single ADU is equivalent to 3.24$\times$10$^{-15}$ erg s$^{-1}$ cm$^{-2}$ above the atmosphere.
To account for the scattering of target photons by the atmosphere of the Earth, it is necessary to modify the expression to include the airmass $X$, becoming $1\ \mathrm{ADU}= 3.24\times10^{-15}\ \mathrm{erg}\ \mathrm{s}^{-1}\ \mathrm{cm}^{-2} \cdot 10^{0.13 X}$. By definition, one rayleigh $R$ is equivalent to $\frac{10^6}{4\pi}$ photons s$^{-1}$ cm$^{-2}$ Sr$^{-1}$. For a photon wavelength of 5890 \AA, this is equivalent to 6.31$\times$10$^{-18}$ erg s$^{-1}$ cm$^{-2}$ arcsec$^{-1}$. Combining these two results leads to the expression
\begin{equation}
    1\ \mathrm{ADU} = 5.12 \cdot 10^{2+0.13X}\ \mathrm{R},
\end{equation}
which allows for the conversion of raw images in ADU to rayleigh.

\subsection{Image background removal}\label{sec:background_removal}

\begin{figure}
    \centering
    \ifthenelse{\equal{\twocol}{"y"}}{
        \captionsetup[subfigure]{width=0.5\linewidth}
        \subfloat[Raw image.]{
            \includegraphics[width=0.52\linewidth]{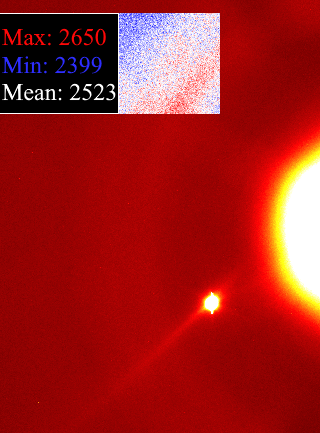}
        } \\
        \subfloat[Extrapolated image background near Io.]{
            \includegraphics[width=0.52\linewidth]{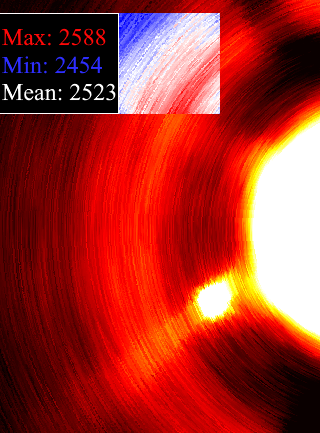}
        } \\
        \subfloat[Image with background near Io removed.]{
            \includegraphics[width=0.52\linewidth]{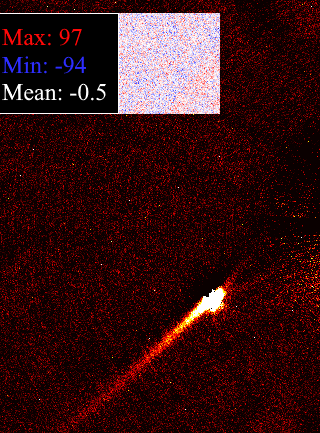}
        }
    }
    {
        \captionsetup[subfigure]{width=0.28\linewidth}
        \subfloat[Raw image.]{
            \includegraphics[width=0.31\linewidth]{BackgroundRemovalExample_orig.png}
        } 
        \subfloat[Extrapolated image background near Io.]{
            \includegraphics[width=0.31\linewidth]{BackgroundRemovalExample_background.png}
        }
        \subfloat[Image with background near Io removed.]{
            \includegraphics[width=0.31\linewidth]{BackgroundRemovalExample_imageMinusBackground.png}
        }
    }
    \caption{
        The results of the background-removal process on a raw image of Io and Jupiter from 2014-12-04. Note that the apparent size of Io is augmented in the images by the overdensity of sodium near the moon. In all images, a square region absent of any bodies has been highlighted to show the effect of the background-removal process on the background pattern. The minimum, maximum, and mean pixel values for this square region have been annotated to one side.
    }
    
    \label{fig:full_image_raw}
\end{figure}

The TRAPPIST telescopes were developed for the imaging of small solar-system bodies (such as asteroids or comets) and for exoplanet detection via the transit method. The use of these telescopes to instead detect the neutral sodium cloud around the relatively bright objects present in the jovian system therefore necessitates additional preprocessing. After image reduction using the dark, bias, and flat frames, there remains a background pattern centred on Jupiter due to the telescope optics, predominantly an internal reflection of light from Jupiter which reveals the shadow of the telescope spider and secondary mirror (see Fig. \ref{fig:full_image_raw}). Once this background has been removed, further preprocessing is required to facilitate the automatic detection of radial structures in the neutral sodium cloud.

To remove the azimuthal background pattern around Jupiter, a region of the image centred on Io is extracted. Since it is observed that the background pattern is approximately azimuthally symmetric about the centre of Jupiter, the image is then projected to an angle-radius projection centred on Jupiter. Gaps in the projected image are linearly interpolated along the azimuthal dimension. To remove it from the cropped image, it is noted that Io is spatially limited and azimuthally asymmetric about the centre of Jupiter, and hence a 10\textdegree\ median filter is applied to the image row at each pixel radius. It was noted a posteriori that the widths of the radial structures identified in this work were not comparable with the spatial extent of Io and the vertical saturation pattern, and so a median filtering with a window width chosen to remove Io from the image is unlikely to affect the strength of the signal from the radial structures. This filtered image is then re-projected into the original image space, with the presence of Io significantly diminished and the azimuthal background pattern still present. The results of this background-removal process are illustrated in Fig. \ref{fig:full_image_raw}b and \ref{fig:full_image_raw}c.

To ensure that it operates as intended, the background-removal process was also applied to a region of an image taken on 2014-12-04 without any bright features, as shown in Fig. \ref{fig:full_image_raw}. The azimuthal pattern present before the removal of the background was greatly diminished and the mean pixel value reduced to approximately zero ADU, which supports the use of this method to remove the azimuthally symmetric background pattern.

\subsection{Artefact removal and preprocessing}\label{sec:preprocessing}

\begin{figure}[t]
    \centering
    \ifthenelse{\equal{\twocol}{"y"}}{
        \captionsetup[subfigure]{width=0.43\linewidth}
        \subfloat[Raw image.]{
            \includegraphics[width=0.45\linewidth]{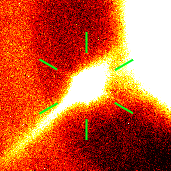}
        }\\
        \subfloat[Comparison with the other moons.]{
            \includegraphics[width=0.45\linewidth]{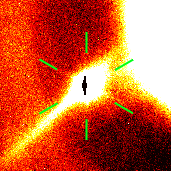}
        }
        \subfloat[Removal of the background.]{
            \includegraphics[width=0.45\linewidth]{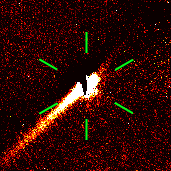}
        }\\
        \subfloat[Removal of the vertical saturation pattern.]{
            \includegraphics[width=0.45\linewidth]{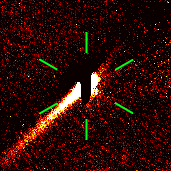}
        }
        \subfloat[Reduction of 60\textdegree\ rotational symmetry.]{
            \includegraphics[width=0.45\linewidth]{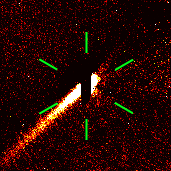}
        }
    }
    {
        \captionsetup[subfigure]{width=0.16\textwidth}
        \subfloat[Raw image.]{
            \includegraphics[width=0.18\textwidth]{Io_crop_raw_diffPatt.png}
            
        }
        \subfloat[Comparison with the other moons.]{
            \includegraphics[width=0.18\textwidth]{Io_crop_MC_diffPatt.png}
        }
        \subfloat[Removal of the background.]{
            \includegraphics[width=0.18\textwidth]{Io_crop_MC_BG_diffPatt.png}
        }
        \subfloat[Removal of the vertical saturation pattern.]{
            \includegraphics[width=0.18\textwidth]{Io_crop_MC_BG_SP_diffPatt.png}
        }
        \subfloat[Reduction of 60\textdegree\ rotational symmetry.]{
            \includegraphics[width=0.18\textwidth]{Io_crop_MC_BG_SP_RR_diffPatt.png}
        }
    }
    \caption{
        Preprocessing steps applied to an image taken on 2014-12-04, cropped around Io. The position of the six-pronged diffraction pattern is denoted by six green lines to aid the eye. It can be seen that the jet becomes more prominent with each step.
    }  
    \label{preprocessing_steps}
\end{figure}

Once the image background close to Io has been removed, it is possible to compare the shape of the other moons present in the image with that of Io, to diminish other image artefacts originating from the telescope. This is performed by cropping the other moons from the image using the same pixel bound as before, normalising and inverting the cropped image, and then multiplying the cropped image of Io by the inverted images of the other moons. Pixels that are bright in the image of Io but dim in the images of the other moons (i.e. sodium-cloud structures) are left unchanged, whereas pixels that are bright in both the image of Io and the images of the other moons (e.g. the diffraction pattern, image artefacts) are diminished. The results of this processing step (as well as the steps detailed below) are given in Fig. \ref{preprocessing_steps}.

The images taken on a particular night undergo the processing described above and are then stacked to increase the signal-to-noise ratio of any structures present. To aid the automated detection algorithm in finding true radial structures in the image data, it becomes necessary to remove or reduce the prominence of both the vertical saturation pattern (most visible in Fig. \ref{fig:full_image_raw}a as a protrusion from the top and bottom of Io) and the six-pronged telescope diffraction pattern (the ``spiderweb''). Additionally, comparing the shape of the image of Io with the other Galilean moons will remove image artefacts that could be erroneously identified as jets. Indeed, no other features that could be interpreted as jets were seen near the other Galilean moons over the course of this work.

To remove the vertical saturation pattern from the image of Io, a suitable pixel bound was identified by taking half of the apparent pixel distance (or a quarter in the case of the far-brighter Jupiter) between Io and the other bodies present in the image, to avoid contamination. Io was then centred in the image and cropped out using this bound. To remove the vertical saturation pattern without diminishing the presence of radial structures in the neutral sodium cloud, it is necessary to remove features with 180\textdegree\ rotational symmetry and vertical reflectional symmetry. By summing the three images (original, 180\textdegree-rotated, vertically flipped) and taking the median, the saturation pattern can be isolated. Subtracting this saturation pattern from the image of Io then leaves true neutral sodium structures untouched whilst removing a source of false detection from the algorithm.

It is possible to leverage its six-fold rotational symmetry to remove the spiderweb diffraction pattern from the images, which may otherwise prove a source of false detections. By rotating the cropped image of Io about its centre in 60\textdegree\ intervals and subtracting this from the original unrotated image, the effect of the spiderweb pattern is dramatically reduced. Whilst the pattern is nevertheless still present in images after this correction (due to an asymmetric diffraction pattern or the error in centring Io in the cropped image), it appears disrupted and hence far less ``jet-like'' for the detection algorithm.

\subsection{Automatic detection of radial structures}

\begin{figure}[t]
    \centering
    \ifthenelse{\equal{\twocol}{"y"}}{
        \includegraphics[width=\linewidth]{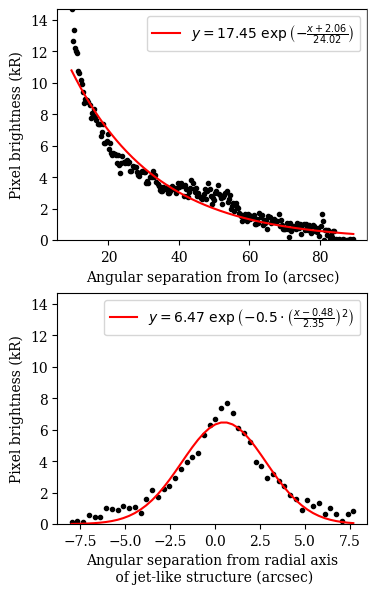}
        \caption{
            Example of fitted radial (top) and azimuthal (bottom) profiles for the jet-like structure observed on 2014-12-04. The fitted profile is given in red against the processed pixel brightnesses in black. 
        }
    }
    {
        \includegraphics[width=\linewidth]{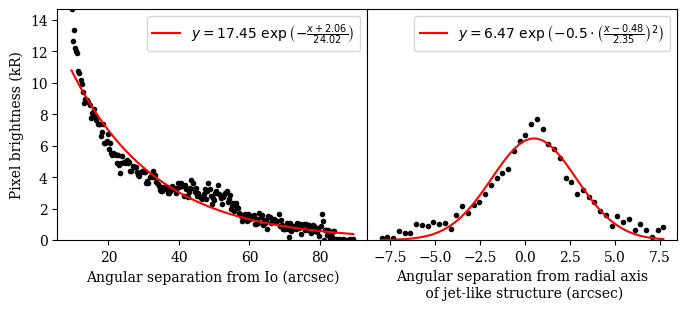}
        \caption{
            Example of fitted radial (left) and azimuthal (right) profiles for the jet-like structure observed on 2014-12-04. The fitted profile is given in red against the processed pixel brightnesses in black. 
        }
    }

    \label{fit}
\end{figure}

Using the preprocessing steps described in the previous sections, an image of Io is obtained in which the saturation pattern and the diffraction pattern are diminished and radial structures highlighted. 
The most visible jets from the first two observation periods (2014-2015 and 2021) allowed for the verification of the radial and azimuthal profiles. As expected from preliminary analysis \citep{spiegeleire2019} and as shown in Fig. \ref{fit}, the radial profile of the jet-like features may be fitted by a decreasing exponential function and the azimuthal profile by a Gaussian function.
Therefore, automatic detection can be carried out by fitting an exponential function to the radial profile and a Gaussian function to the azimuthal profile at various intervals around Io. 

Taking the central angle in 1\textdegree\ intervals for the full 360\textdegree\ around Io, a 10\textdegree\ region, centred on the central angle, is evaluated for the presence of a radial structure. 
This region is summed along the azimuthal axis, then normalised. A decreasing exponential function of the form ${y=S_y e^{-x/S_x}}$, where $x$ is the radial distance from Io in arcseconds, $y$ is the pixel brightness in kR, and ${S_x}$ and ${S_y}$ are constants to be found, is then fitted to the radial profile, and the R${^2}$ goodness of fit evaluated for the fitting. 
Gaussian profiles of the form ${y=a\exp{(-\frac{x^2}{2\sigma^2}})}$, where ${a}$ and ${\sigma}$ are again constants to be found, are fitted at ten evenly spaced radial points between a lower radial limit (15 pixels from the centre of Io; chosen to avoid contamination of the jet from sunlight reflected by Io) and the upper pixel bound identified previously, and the R${^2}$ goodness of fit again evaluated for each of these Gaussian profiles. 
This method is preferred over a summation of the image over the radial dimension as it ensures that the azimuthal profile of the structure is indeed Gaussian at a range of distances from Io.  
The median of the R${^2}$ values is used as a measure of the goodness of fit of an azimuthal Gaussian profile to the structure. 
The product of the azimuthal and radial goodness-of-fit measures is taken as the indicator of the likelihood that a jet-like structure be present at this central angle (the ``jet value''), which ranges from 0 (profile at this central angle poorly described by a jet) to 1 (profile at this central angle well-described by a jet). 

It is worth noting that this method merely provides the angles around Io which best show a jet-like profile; it remains for the user to decide whether the angles returned show sufficiently jet-like appearances to be reasonably interpreted as jet-like structures in the neutral sodium cloud. To this end, the algorithm returns the cropped image of Io with suitable lower and upper brightness bounds such as to maximise the variation in brightness within the segment of image around the detected jet-like angle(s). If no structure is visible even with these ideal brightness bounds, it is reasonable to conclude that no structure is present for this date. If a structure appears to be present at the detected angle, the processed image of Io is compared with processed images of the other moons; if the same jet-like structure is observed around another moon, it is assumed to be a telescope artefact or related to the diffraction pattern, and the structure candidate discarded. While Europa also has a neutral sodium cloud \citep{burgerandjohnson2004}, the presence of jet-like features is yet to be reported in the literature and, even if present, a jet-like feature in the neutral sodium cloud of Europa would be unlikely to have the same observed instantaneous orientation as the feature in the cloud of Io.

\begin{figure}[h]
    \centering
    \ifthenelse{\equal{\twocol}{"y"}}{
        \includegraphics[width=\linewidth]{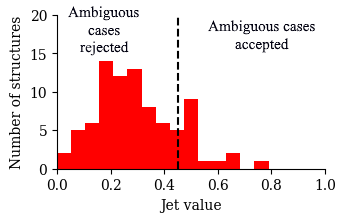}
    }
    {
        \includegraphics[width=\linewidth]{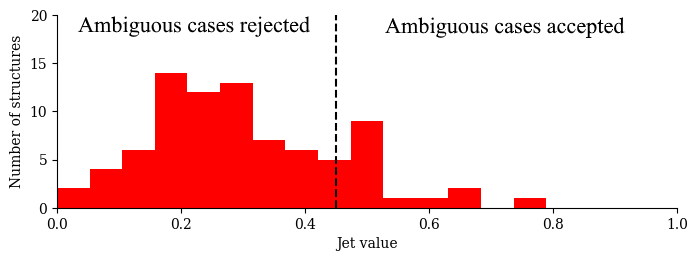}
    }
    \caption{
        Histograms of the jet value of all jet-like-structure candidates, with a bin width of 0.05. A dashed line has been annotated to indicate the jet-value cutoff for ambiguous cases.
    }  
    \label{fig:jet_value_histogram}
\end{figure}

In order to establish an objective threshold between true and false positive results from the auto-detection process, the distribution of jet value over all auto-detected structures was investigated; see Fig. \ref{fig:jet_value_histogram}. The jet values of the structures show a broadly bimodal distribution surrounding a jet value of 0.45. This limiting jet value was therefore taken as a stable cutoff to distinguish between true and false positives. However, this cutoff would lead to the discarding of several cases in which a clear radial structure is observed (due to noise or imperfect detection of the position of Io in the image); the results of this cutoff operation were therefore checked by a human operator and these misattributed cases nevertheless classed as positive detections of a radial structure in the neutral sodium cloud.

\subsection{Magnetic-field model}

This work uses the JRM33 internal-magnetic-field model of Jupiter \citep{connerneyEA2022} in conjunction with the Con2020 model of the external magnetic field due to the equatorial current sheet \citep{connerneyEA2020} to model the magnetic field close to Io. These models are accessed via the JupiterMag Python wrapper made available as part of the Magnetospheres of the Outer Planets Community Code project \citep{jupitermag}.

\section{Results}

It is first necessary to identify the observational characteristics of the different structures in the Io neutral sodium cloud, primarily those of jets and of the banana, to allow for the interpretation of the images presented in this work.
\begin{itemize}
    \item Jets are presumed to line approximately in the plane perpendicular to the local magnetic field at Io and to extend exclusively in the anti-jovian direction \citep{wilsonEA2002}. Therefore, structures displaying a jet-like morphology that extend in an anti-jovian direction when projected to the plane perpendicular to the local magnetic field at Io can be reliably interpreted as jets.
    \item The neutral sodium banana may appear observationally similar to jets in the sodium cloud. However, the banana is aligned with the orbital plane of Io, rather than the plane perpendicular to the local magnetic field, and is directed along, or slightly internal to, the orbit of Io \citep{wilsonEA2002}. Therefore, structures with a jet-like morphology that appear to extend inside the orbit of Io when projected to the plane perpendicular to the local magnetic field, and that are well aligned with the apparent direction of movement of Io, may be interpreted as observations of the banana rather than of a jet. 
\end{itemize}

Images of Io processed according to Sects. \ref{sec:background_removal} and \ref{sec:preprocessing} for all observing runs discussed in this work are given in Fig. S1 in the supplementary material.

\begin{figure*}[t]
    \centering

    \subfloat[2014-12-04]{
        \includegraphics[width=\linewidth]{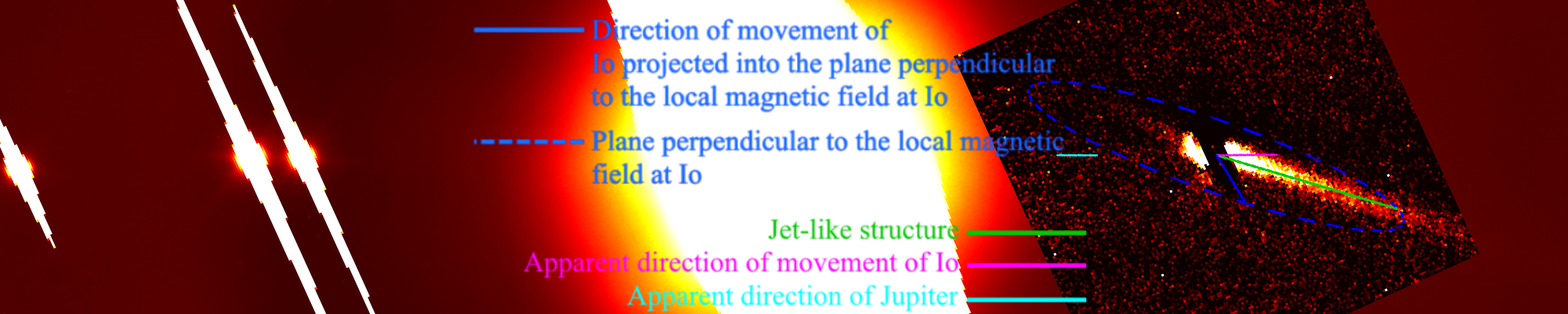}
    }
    \hspace{2pt}
    \subfloat[2015-01-28]{
        \includegraphics[width=\linewidth]{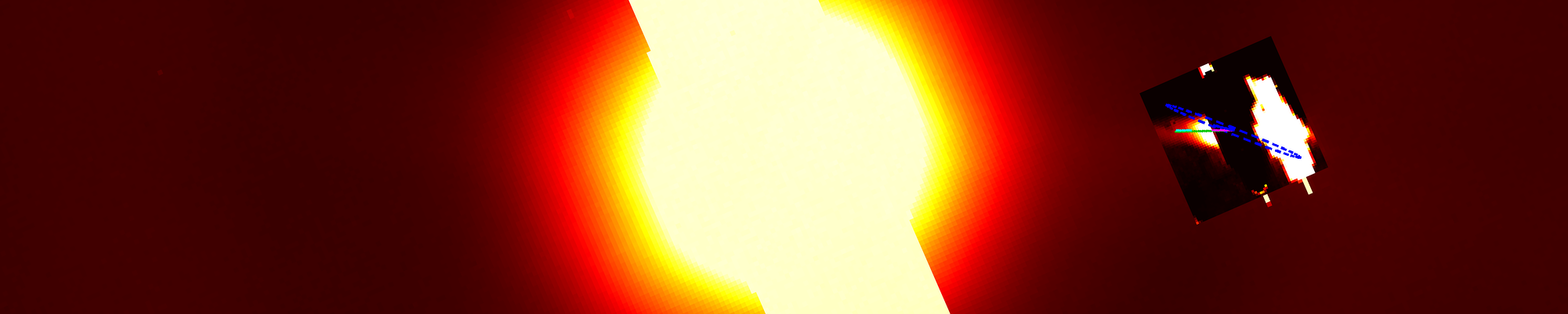}
    }
    \\
    \subfloat[2015-02-12]{
        \includegraphics[width=\linewidth]{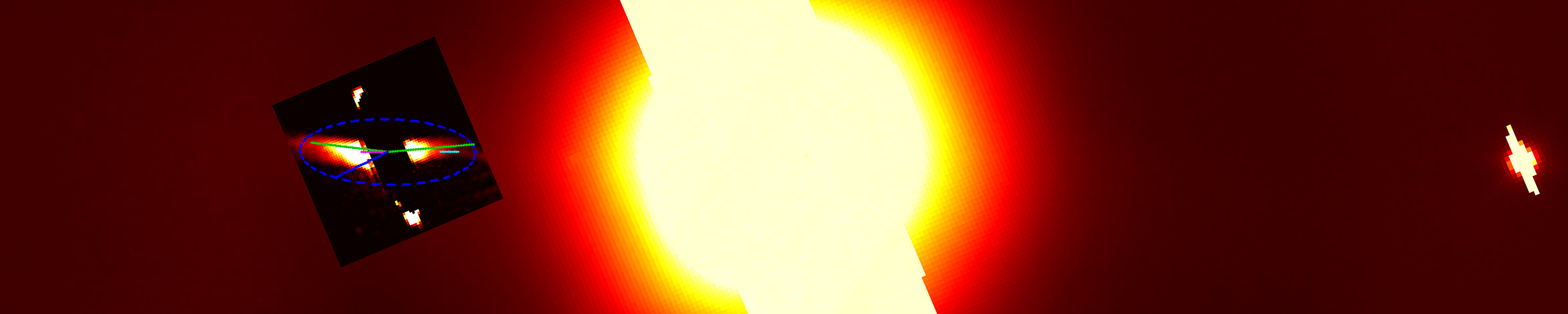}
    }
    \\
    \subfloat[2022-07-03]{
        \includegraphics[width=\linewidth]{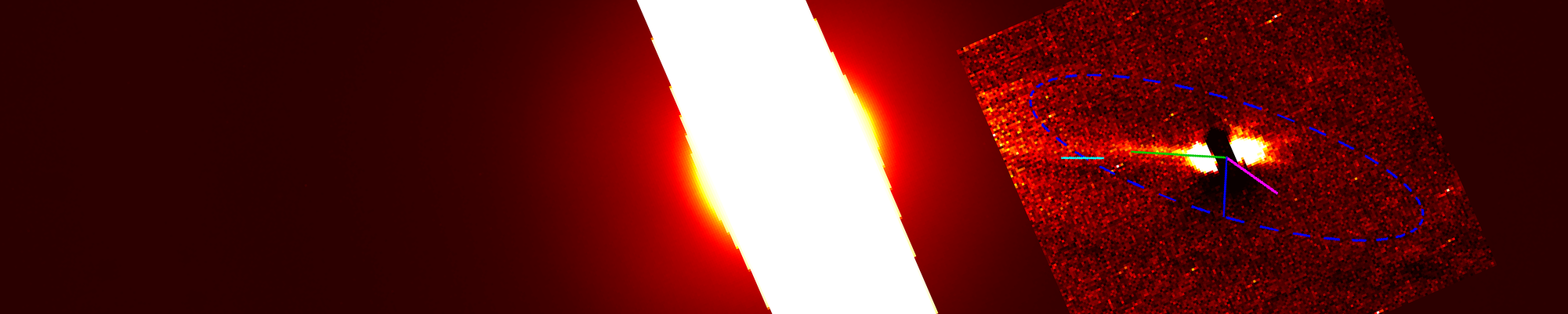}
    }

    \caption{
        A sample of results of the automatic jet detection, overlaid on images of Io processed as per Sects. \ref{sec:background_removal} and \ref{sec:preprocessing}. Images are orientated with north upward and west to the right, and Jupiter is located in the centre. The direction toward the centre of Jupiter in the image is indicated by a short cyan line. The short magenta line centred at Io indicates the apparent direction of movement of Io in the image. The dashed blue ellipse represents the plane perpendicular to the local magnetic field at Io according to an observer on Earth, and the blue line between the centre of Io and the edge of this ellipse is the projection of the movement vector of Io in this plane. Green lines represent detected jet-like structures.
    } 
    \label{fig:jets_detected}
\end{figure*}

The imaging campaign of 2014-2015, despite being intended as a test of the ability of the TRAPPIST telescopes to observe the neutral sodium cloud, produced several clear images of jet-like structures. Of the 17 nights on which observations of Io took place, a jet-like structure was present for 10 of them. The structures detected vary in length and brightness, often changing their appearance greatly between consecutive nights.

In particular, the structure detected on 2014-12-04 (see Fig. \ref{fig:jets_detected}a) is notable for both its length and brightness compared to all other structures found in this work. 
This structure, if presumed to be in the plane perpendicular to the local magnetic field, extended in the anti-jovian direction (+78\textdegree\ from the projected direction of movement of Io) and is thus readily interpreted as a jet.
Though the banana cloud is typically the most prominent of the structures in the neutral sodium cloud \citep{gravaEA2021}, this jet is far clearer than any other structure in this image.
This jet is also remarkable for its apparent thinness, which may imply that sodium particles are ejected with single-value launch speeds or angles.
This is in disagreement with the "fan-like" structure presented in e.g. \citet{wilsonEA2002}, which would result in a thicker jet, especially further from Io.
A similarly clear (though visibly diminished) jet was also detected on 2014-12-06, which may be a continuation of the jet of 2014-12-04. 
The observation of 2014-12-08 is hampered by the proximity of Io to Jupiter, and the observation of 2014-12-12 by the proximity of Io to Europa, which have prevented the detection of the jet on the nights, if still present. 

The jet detected on 2014-12-04 appears very similar to the observation of the "stream" neutral cloud from 1990-01-12 discussed by \citet{schneiderEA1991}. 
Both structures show considerable lengths, are limited in width, and are directed away from Jupiter in the plane perpendicular to the magnetic field.
However, the observation of the stream by \citet{schneiderEA1991} showed a "hook" at its most distant point which is not present in the jet of 2014-12-04, despite the comparable lengths of the two structures.
Nevertheless, the similarity between these two structures highlights the difficulty in visually distinguishing between the jet and a stream originated from the plasma torus close to Io, as the two are produced via very similar mechanisms and differ only in the duration between pickup and reneutralisation of the sodium \citep{wilsonEA2002}.
In the viewing configuration of 1990-01-12, an anti-jovian jet that is emitted within the plane perpendicular to the local magnetic field would be masked by the stream, and hence it is not possible to say whether the stream and the jet are genuinely the same structure in this case, or whether the jet (if present) is simply not visible.

\begin{figure}[t]
    \centering
    \ifthenelse{\equal{\twocol}{"y"}}{
        \subfloat[2014-2015]{
            \includegraphics[width=0.6\linewidth]{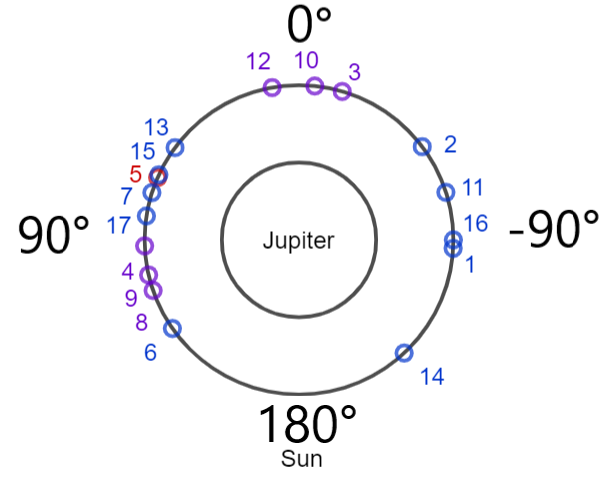}
        }\\
        \subfloat[2021]{
            \includegraphics[width=0.6\linewidth]{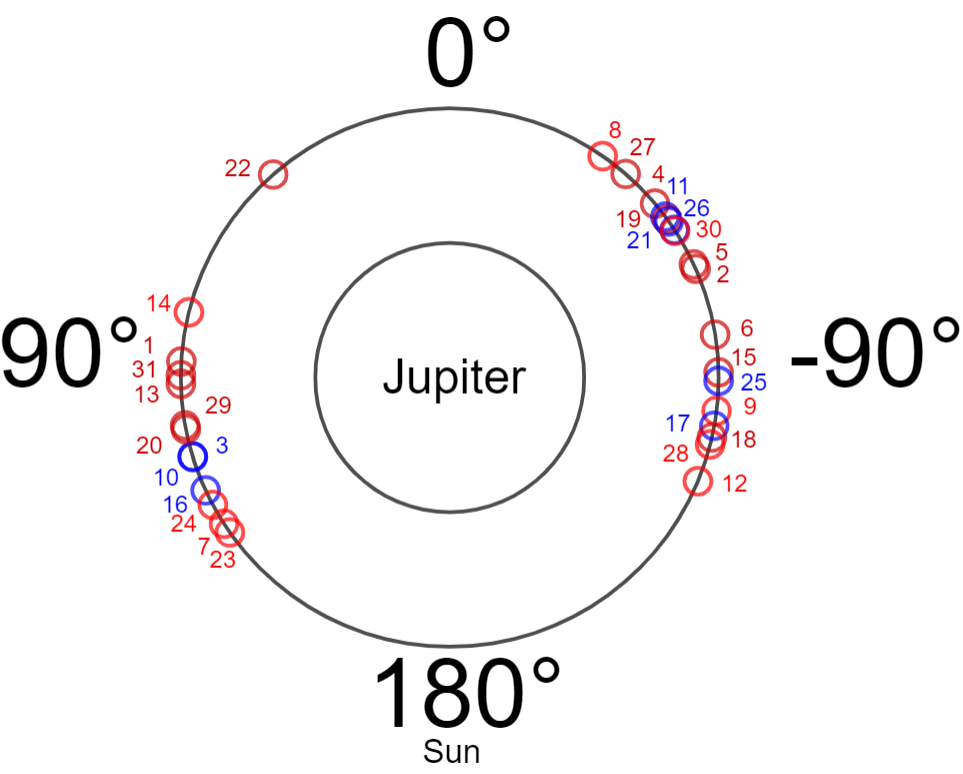}
        }\\
        \subfloat[2022-2023]{
            \includegraphics[width=0.6\linewidth]{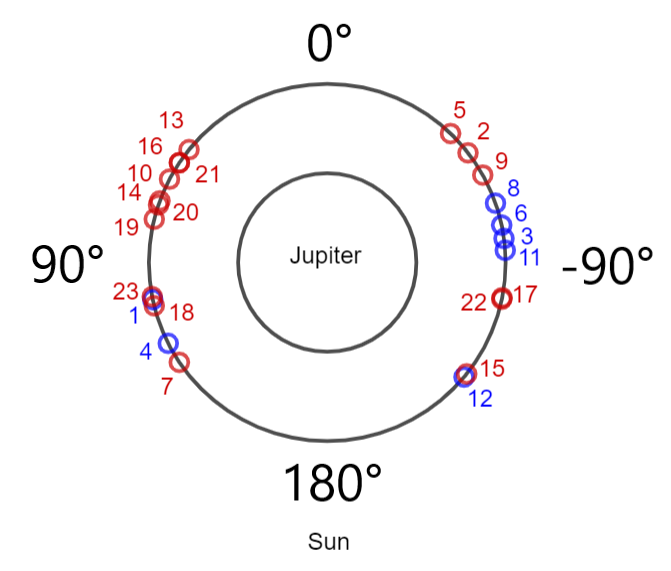}
        }
    }
    {
        \subfloat[2014-2015]{
            \includegraphics[width=0.3333\textwidth]{2014-2015_Sun.png}
        }
        \subfloat[2021]{
            \includegraphics[width=0.3333\textwidth]{Sun_2021_geo_fin4D.png}
        }
        \subfloat[2022-2023]{
            \includegraphics[width=0.3333\textwidth]{2022-2023_SUN.png}
        }
    }
    \caption{
        Io phase angle relative to the Sun for the observations detailed in this work. The position of Io is given as a small circle where the corresponding number refers to the case index given in tables \ref{tab:obs_2014}, \ref{tab:obs_2021}, and \ref{tab:obs_2022}. Red circles indicate that no jet-like structure was observed on this date, whereas a blue circle indicates that at least one jet-like structure was observed. The cases in purple are those for which Io is behind Jupiter or close to another moon, so non-detection during these observations may be due to the configuration of the system as well as the absence or the faintness of the jet itself. Key phase angles, given in degrees, have been annotated around the diagram, as well as the position of the Sun in the diagram.
    }
    \label{fig:phase_angle_sun}
\end{figure}

\begin{figure}
    \centering
    \ifthenelse{\equal{\twocol}{"y"}}{
        \subfloat[2014-2015]{
            \includegraphics[width=0.60\linewidth]{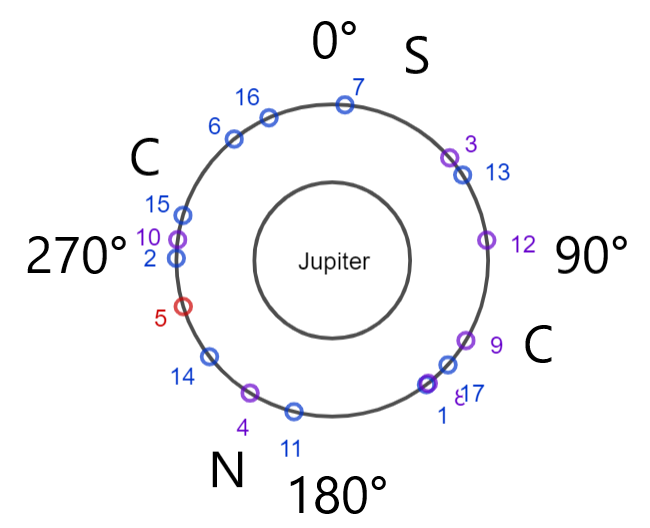}
        }\\
        \subfloat[2021]{
            \includegraphics[width=0.60\linewidth]{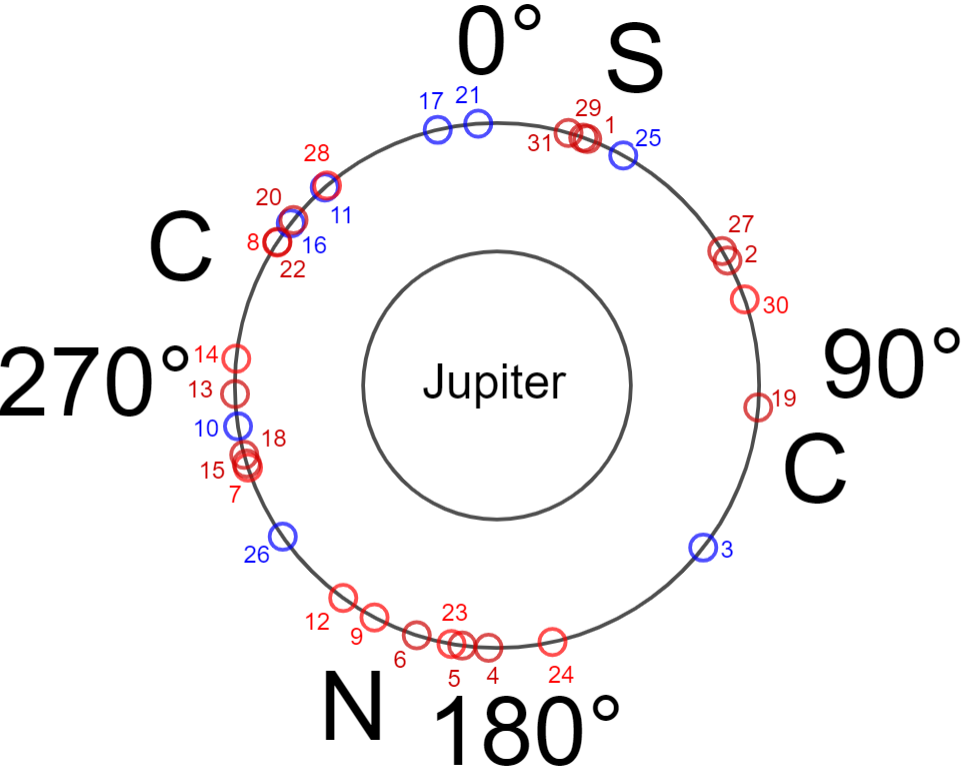}
        }\\
        \subfloat[2022-2023]{
            \includegraphics[width=0.60\linewidth]{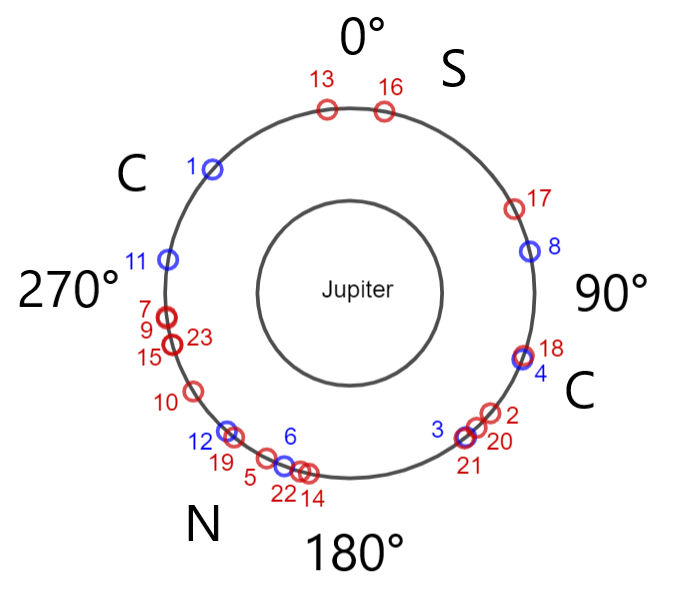}
        }
    }
    {
        \subfloat[2014-2015]{
            \includegraphics[width=0.3333\textwidth]{2014-2015_S3.png}
        }
        \subfloat[2021]{
            \includegraphics[width=0.3333\textwidth]{S3_2021_geo_fin4D.png}
        }
        \subfloat[2022-2023]{
            \includegraphics[width=0.3333\textwidth]{2022-2023_S3.png}
        }
    }
    \caption{
        Position of Io in System-III longitude for the observations detailed in this work. The annotation of the positions of Io are as in Fig. \ref{fig:phase_angle_sun}. The location of Io in the plasma torus is annotated, where ``N'', ``S'', and ``C'' indicate that Io is located northward, southward, or in the centre of the plasma torus. Key System-III longitudes, given in degrees, have been annotated around the diagram.
    }
    \label{fig:system3}
\end{figure}

In 2021, while jet-like structures are again observed in the images from the TRAPPIST telescopes, they are noticeably less distinct than those structures observed in 2014. 
This variation cannot be uniquely due to the phase angle nor the System-III longitude of Io, since both parameters are well sampled in both the 2014-2015 and the 2021 campaigns, as shown in Figs. \ref{fig:phase_angle_sun} and \ref{fig:system3}. 
Indeed, a case-by-case examination returns cases with very similar phase angles (e.g. 2014-12-21 and 2021-07-08; see Fig. S2) or very similar System-III longitudes (e.g. 2014-12-06 and 2021-05-31; see Fig. S3) in which a jet-like structure is visible in one set of images and not in the other. 
The 2021 campaign also demonstrated that the variation of the jet may be present over a timescale as short as a day. As shown in table \ref{tab:obs_2021}, no jet was detected on the night of 2021-06-23, whereas a jet appeared during the observation on the following night, 2021-06-24 (see Fig. S4). 

In the 2022-2023 imaging campaign, the jet-like structures, of which several were observed, are again less distinct than in the 2014-2015 campaign. 
Compared to the very clear jet observed on 2014-12-04, cases without jet-like structures were encountered in the 2022-2023 campaign despite similar phase angle (e.g. 2022-12-23) or System-III longitude (e.g. 2023-01-14). 
Indeed, both of these parameters were very similar on 2022-12-22 ($\phi_{E}$ = 67\textdegree, $\theta_{S3}$ = 11\textdegree) and on 2015-01-11 ($\phi_{E}$ = 68\textdegree, $\theta_{S3}$ = 5\textdegree); however, as shown in Fig. S5, while a clear jet-like structure was observed on this latter date, the same orbital configuration did not produce any jet-like structures for the observation in 2022.
These results imply that neither phase angle nor System-III longitude nor a combination of the two are uniquely responsible for the presence or absence of jet, and that this must instead by largely controlled by some other process.

In several cases, a jet-like structure was observed which extended toward Jupiter, such as 2015-01-28 (see Fig. \ref{fig:jets_detected}b). 
In this case, the structure cannot be readily interpreted as a jet.
However, due to the excellent alignment of this structure with both the apparent Io-Jupiter direction and the apparent direction of movement of Io, it may instead be interpreted as a detection of the banana neutral sodium cloud.

On several nights, such as 2015-02-12 (see Fig. \ref{fig:jets_detected}c), multiple jet-like structures were observed. 
In this case, one structure is well-aligned with the apparent direction of movement of Io and anti-aligned with the location of Jupiter on the image, which is the expected behaviour for the banana neutral sodium cloud in this configuration. 
The other structure, when projected to the plane perpendicular to the local magnetic field at Io, is directed toward Jupiter (-123\textdegree\ from the projected direction of movement of Io, though the small tilt of this plane makes exact determination of this angle difficult) but is anti-aligned with the apparent direction of movement of Io. 
Thus, it is not readily identifiable as either a jet or the banana. 
However, the banana does show a slight curvature inward of the orbit of Io in the jovian direction \citep{wilsonEA2002}, which, combined with the short observed length of the structure and the fact that Io was observed to be moving largely perpendicular to the viewing plane (phase angle 55\textdegree), implies that this case may be another detection of the banana neutral sodium cloud.

Jet-like structures that extended toward Jupiter but not in the apparent direction of movement of Io  were also detected in several cases during this campaign; see Fig. \ref{fig:jets_detected}d for an example.
It cannot be stated with certainty whether this structure represents a jet or the banana or indeed another structure in the neutral sodium cloud.
It is possible that this is another detection of the inward-curving geometry of the banana, or is simply a result of the image artefacts that can be seen to the left and right of Io.

\begin{figure}
    \centering
    \ifthenelse{\equal{\twocol}{"y"}}{
        \includegraphics[width = \linewidth]{   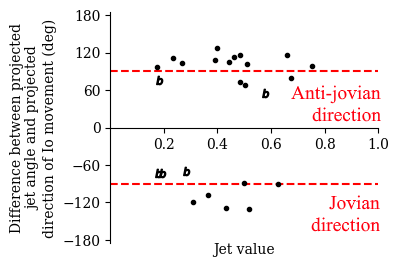}
    }
    {
        \includegraphics[width = \textwidth]{   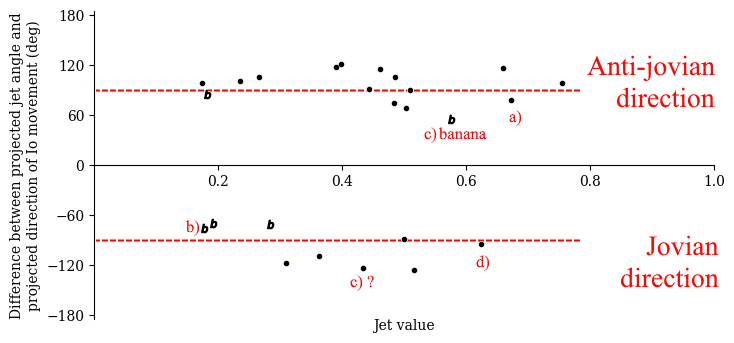}
    }
    \caption{
        Jet value of detected structures against the angular deviation from the direction of movement of Io projected into the plane perpendicular to the local magnetic field at Io. Angles greater than 0\textdegree\ indicate an anti-jovian direction, whereas angles smaller than 0\textdegree\ indicate a jovian direction (inside the orbit of Io). Cases marked with a ``b'' are those for which the structure is well-aligned with the expected direction of the banana. Dashed lines indicating the expected orientation of an exactly anti-jovian (+90\textdegree) and jovian (-90\textdegree) structure have been included to guide the eye. Annotation has been applied below those cases displayed in Fig. \ref{fig:jets_detected}.
    } 
    \label{fig:angular_deviation}
\end{figure}

In this study, besides the cases explored above, a variety of structures were detected in the Io neutral sodium cloud with a variety of angular deviations from the movement vector of Io projected into the plane perpendicular to the local magnetic field, as shown in Fig. \ref{fig:angular_deviation}. 
The majority of these structures extended in the anti-jovian direction, including many structures with high jet values (well described by jet-like profiles), and can therefore be readily interpreted as jets in the neutral sodium cloud.
These structures are grouped around a projection angle of +90\textdegree\ from the projected movement direction of Io in the plane perpendicular to the local magnetic field (i.e. almost exactly anti-jovian). This may be a selection effect rather than a physical preference; the vast majority of the observations discussed in this work were made when Io was moving toward or away from the observer, and so any structure aligned with the direction of movement of Io would be hidden by the emission from Io itself.  
Several structures that extended in the jovian direction also demonstrated the expected behaviour of the banana neutral sodium cloud.
Of the remaining structures extending in the jovian direction detected in this work, many can be explained by one or more of the following restrictions:
\begin{itemize}
    \item The non-uniform morphology of the banana and the degree of movement of Io perpendicular to the viewing plane may have led to a detection of the banana that was anti-aligned to the apparent direction of movement of Io;
    \item Io was observed close to Jupiter and the observed structure is well aligned with the apparent position of Jupiter in the image. Here, it is possible that the background-removal process and proximity to Jupiter is causing a jet-like artefact to appear in the images;
    \item The detected structure may be an artefact of the observation process, such as a limb of the saturation pattern not fully removed by preprocessing; here, the detected structure may be well aligned with the saturation pattern (though still not present in the processed images of the other moons, otherwise it would have been discarded after the auto-detection) or be significantly dimmer compared to other, clearer structures.
\end{itemize}
While this does not preclude the possibility that these cases be legitimate detections of a structure in the neutral sodium cloud that is neither a jet nor the banana, further observation is required to provide a detection that cannot be explained by the above restrictions.
Nevertheless, there remain cases, such as that shown in Fig. \ref{fig:jets_detected}c, where a clear structure is observed to be directed toward Jupiter and away from the apparent direction of movement of Io. 
We propose that these cases may be detections of the sputtering of sodium from sodium-bearing molecules in dust grains within the orbit of Io \citep{gravaEA2021}.

\begin{table*}[t]
    \caption{
       Derived parameters for the jets identified in this work. ``Length'' refers to the extrapolated intrinsic length of the jet, assuming that it lies in the plane perpendicular to the local magnetic field at Io.
    }
    \begin{center}

        \begin{tabular}{| c | c | c | c | c |}

            \hline
            Date & Brightness at 70\,000 km (kR) & Length (km) & Length (R$_{Io}$) & Length (R$_J$)\\
            \hline
            2014-12-04 & 8.42  & 184\,000 & 101   & 2.6 \\
            2014-12-06 & 1.87  & 63\,000 & 35    & 0.9 \\
            2014-12-21 & 0.76  & 66\,000 & 36    & 0.9 \\
            2015-01-11 & 2.03  & 62\,000 & 34    & 0.9 \\
            2015-02-12 & 3.10  & 67\,000 & 37    & 0.9 \\
            2015-03-30 & 1.87  & 64\,000 & 35    & 0.9 \\
            2015-03-31 & 1.14  & 82\,000 & 45    & 1.1 \\
            2021-05-13 & 1.55  & 79\,000 & 43    & 1.1 \\
            2021-06-06 & 1.07  & 101\,000 & 55    & 1.4 \\
            2021-06-23 & 5.54  & 81\,000 & 44    & 1.1 \\
            2021-07-15 & 2.72  & 54\,000 & 30    & 0.8 \\
            2021-07-22 & 0.84  & 63\,000 & 35    & 0.9 \\
            2022-05-26 & 1.08  & 83\,000 & 46    & 1.2 \\
            2022-09-03 & 2.97  & 48\,000 & 26    & 0.7 \\
            2022-09-05 & 2.42  & 65\,000 & 36    & 0.9 \\
            2022-11-23 & 3.20  & 58\,000 & 32    & 0.8 \\

            \hline
        \end{tabular}
    \end{center}
    \label{tab:derived_parametres}
\end{table*}

The apparent length of the jets identified in this work can be estimated by using the fitted radial exponential profiles. 
To ensure consistency between cases, the distance from Io at which the fitted profile descends below 10\% of the average jet pixel value at the inner detection radius of 15 px was taken as a representative measure of structure length. 
Since it has been assumed that a jet will lie in the plane perpendicular to the local magnetic field at Io, it is possible to infer an intrinsic length from the apparent length and viewing geometry; derived intrinsic lengths for the jets identified in this work are given in table \ref{tab:derived_parametres}.
Using this consistent measure, it can be seen that jets are observed with a large range of intrinsic lengths, which can extend up to several hundred Io radii in the case of the clear jet of 2014-12-04.

Similarly, by removing the background from the reduced images and converting to rayleigh as per Sect. \ref{sec:brightness_conversion}, it is possible to compare the absolute brightness of the detected jets in a consistent manner. To allow for comparison between different cases, the average brightness in a 10\textdegree\ arc about the central axis of the jet at an apparent distance of 70\,000 km from the centre of Io was taken as a representative measure of jet brightness. This is distant enough from Io to be unaffected by reflected sunlight while remaining close enough that the jets present in this work are still detectable. The jet brightnesses obtained using this method are given in table \ref{tab:derived_parametres}. The range of brightnesses observed falls within the expected 1 - 10 kR range for structures in the neutral sodium cloud of Io \citep{smyth1992}. The clear jet of 2014-12-04 is almost twice as bright as any other detected structure.

The jets identified in this work do not show the expected "fan-like" or "hook" shape \citep{wilsonEA2002} but instead present themselves as thin, collimated structures, especially in the case of the jet of 2014-12-04 (Fig. \ref{fig:jets_detected}a). 
While the length of the jet is related to the movement of reneutralised particles along their former magnetic gyrorotation axes, the jet width originates from the former movement of these particles parallel to the magnetic field at Io.
Thus, a collimated jet implies a lack of a considerable component of velocity of pickup ions parallel to the magnetic field.
This may have two explanations.
Firstly, the neutral atoms that become ionised via pickup ionisation are very quickly reneutralised and ejected in a jet.
This would not give the pickup ions sufficient time to attain thermal equilibrium parallel to the magnetic field, since they are generated from Io's relatively cold atmosphere \citep{lellouchEA2007} and hence does not start with a large parallel velocity component. 
Otherwise, it is possible that the pickup-ion plasma itself remains cold parallel to the magnetic field. 
In this case, the parallel temperature of the pickup ions can be estimated from the ratio of jet width to jet length, which gives a maximum launch angle from the central axis for particles in the jet.
For the jet of 2014-12-04, if particles are assumed to have a velocity along the jet axis of 100 km s$^{-1}$ \citep{bagenalanddols2020}, the ratio between the length (126") and the width (20"), calculated from the distance from or along the central axis at which the pixel value descends below 10\% of the peak pixel value, can be taken to arrive at a parallel temperature of sodium pickup ions of 7 eV.
This would imply that the pickup-ion plasma remains relatively cold and consistent with the expected temperatures of neutralised pickup ions escaping from Io \citep{bagenalanddols2020}. A more in-depth analysis using modelling tools would further constrain this result.

\begin{figure}
    \centering
    \ifthenelse{\equal{\twocol}{"y"}}{
        \includegraphics[width=\linewidth]{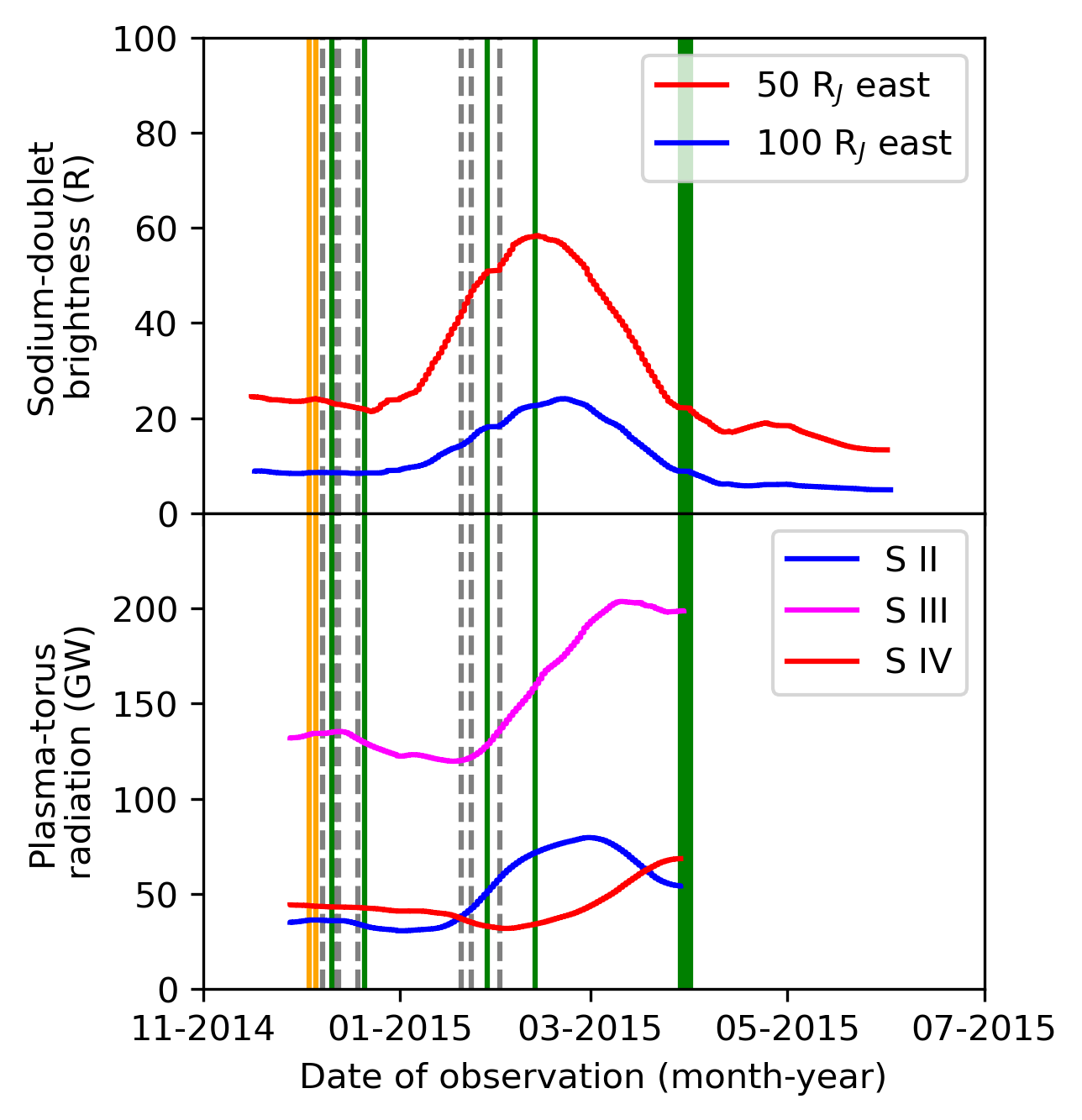}
    }
    {
         \includegraphics[width=\linewidth]{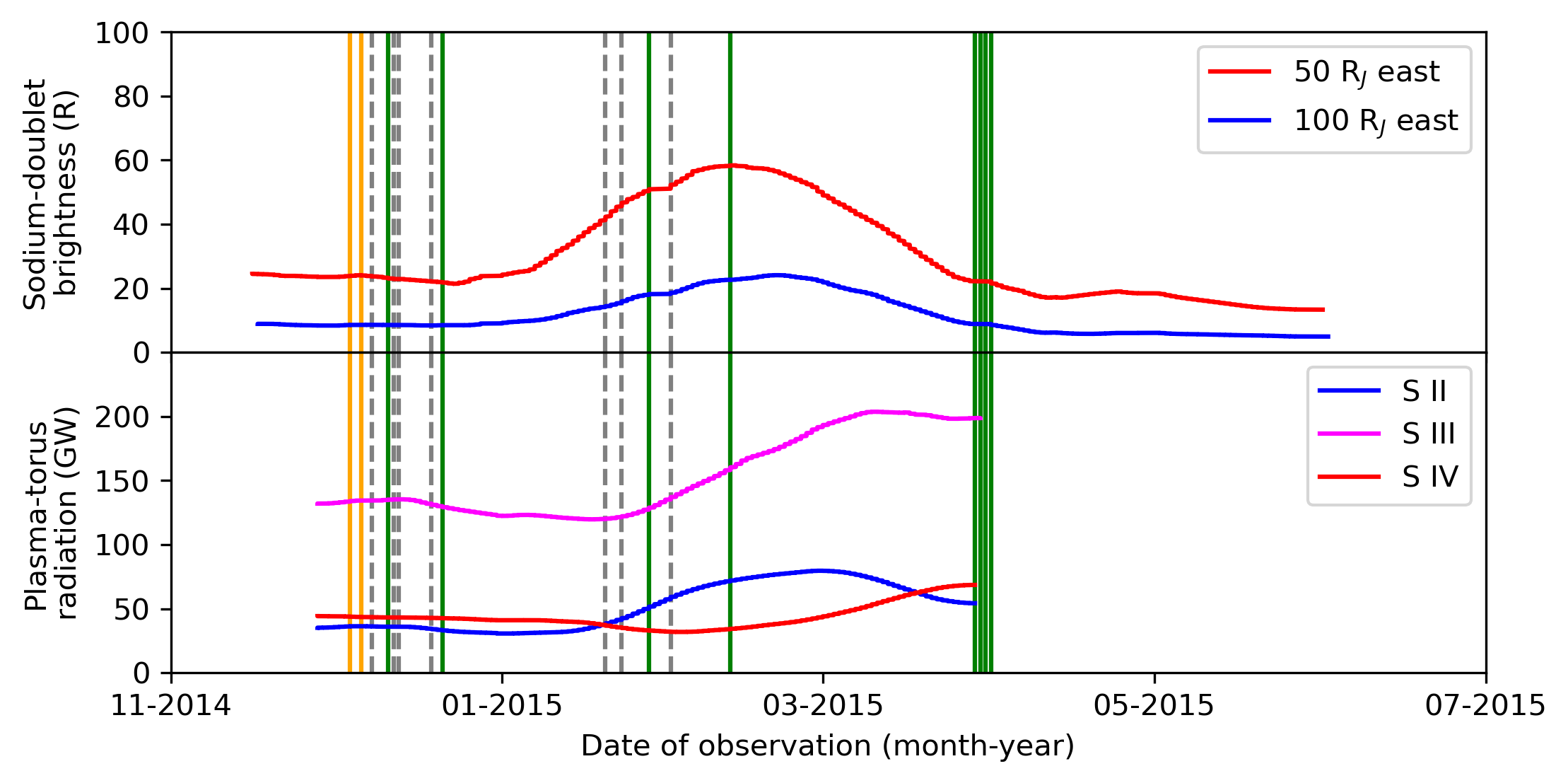}
    }
    
    \caption{
        Comparison between the detections of jets during TRAPPIST observations with previous studies of the brightness of components of the jovian magnetosphere for the period surrounding the 2014-2015 viewing campaign. Data has been extracted from figures in the reference works. Jets identified in this work have been annotated by solid green lines, and observations made of Io with no detected jets have been annotated by a broken grey line. Orange lines denote the exceptional jets of 2014-12-04 and 2014-12-06.\\Top: the brightness of the extended sodium nebula in the sodium-doublet waveband, as taken from Fig. 1 of \citet{yonedaEA2015}. The legend refers to the distance from Jupiter at which the measurements were obtained.\\Bottom: the brightness of the Io plasma torus in several sulphur-ion wavebands, as taken from Fig. 3 of \citet{yoshikawaEA2017}. The legend refers to the ionisation levels of the sulphur ions.
    }
    \label{fig:yoneda-yoshikawa_figure}
\end{figure}

As shown in Fig. \ref{fig:yoneda-yoshikawa_figure}, the brightness of both the extended sodium nebula \citep{yonedaEA2015} and the plasma torus \citep{yoshikawaEA2017} increased towards the end of January 2015. In the period preceding this increase in brightness, multiple jets were observed with the TRAPPIST telescopes. Since the brightness of both the extended sodium nebula and the plasma torus is not observed to increase with the presence or absence of these jets, we conclude that a single instance of a jet does not considerably alter the brightness of these structures. This conclusion is reinforced by the lack of response even to the bright jet of 2014-04-12 (the first blue line in Fig. \ref{fig:yoneda-yoshikawa_figure}).

It is possible that the jets observed in the 2014-2015 campaign are simply one long-lasting structure, albeit of varying length and brightness. While there were nights during this campaign on which no jet was observed, this may have been due to a variable intrinsic brightness or unfavourable viewing geometry that rendered the jet undetectable by the TRAPPIST telescopes. However, were this the case, it remains to be explained why the increase in brightness of the sodium nebula and plasma torus begins much later than the first observed jet, which itself may have arisen before the start of the 2014-2015 viewing campaign. The speed of the neutral sodium ejected by a jet is such that a distance of 100 R$_J$ could be achieved in 24 hours \citep{yonedaEA2015}, and hence any response from the plasma torus or sodium nebula should be rapidly observable. If there is indeed a link between neutral sodium jets and the brightness of the plasma torus or sodium nebula, this result implies a more complex process than a simple input of matter.

\section{Conclusion}

The regular, long-term monitoring of the neutral sodium cloud of Io performed for the first time with the TRAPPIST telescopes led to a total of 25 detections of jet-like structures, with a particularly spectacular case on 2014-12-04, and the physical properties of these jet-like structures were established. Even if the number of detections does not allow for the determination of the precise time variation of those properties, by comparing observations made on different nights, it can be determined that the presence, length, and brightness of jet-like structures do not, or not only, depend on the orbital angle nor the System-III longitude of Io. Additionally, a jet can be clearly present during one night and entirely absent during the next, or the length and brightness of a jet observed over two or more consecutive nights can vary considerably within these observations. Future work should be considered to determine which physical processes most closely control the appearance of jets.

The geometry of a large proportion of the detected jet-like structures aligned well with the expected geometry of sodium jets. Additionally, five cases were observed in which a jet-like structure could be clearly associated with the banana neutral sodium cloud. Many cases which did not fall into these two categories could be explained by the alignment inward of the orbit of Io of the principal axis of the banana. Of the remaining cases, which are observed to extend in an jovian direction but which cannot be explained by the banana, there is no unequivocal detection of a jet-like structure. While this does not preclude the presence of an unexplained jet-like structure in these cases, a clearer detection would be necessary to make firm conclusions.

A comparison between our data and the data from \citet{yoshikawaEA2017} and \citet{yonedaEA2015} shows that the relation between the presence of a jet and the brightness of larger structures, such as the extended sodium nebula or the plasma torus, is not straightforward. Even a bright jet, such as that of 2014-12-04, did not led to an immediate increase of brightness in the plasma torus and extended nebula, and the increase in brightness observed in these larger structures toward the end of January 2015 was not preceded by an especially large jet. This may imply that jets do not contribute directly to the population of these structures.

This work presents a database of jet detections spanning three periods between 2014 and 2023, with many observations made within each period. It is hoped that future work in neighbouring fields may make use of this database to probe the relationship between sodium jets and potentially related datasets, such as those pertaining to volcanism on Io, the other neutral sodium clouds, or the aurorae of Jupiter, in the same way as this work has compared this database to the brightness of the plasma torus and extended sodium nebula. The planned continued regular monitoring of the Io neutral sodium cloud with the TRAPPIST telescopes will serve to enhance the value of this database.

This work also highlighted the need for modelling tools to explain the lack of fan-like structure in the jets identified in this work, as well as to more rigorously derive characteristics of the Io pickup-ion plasma from the observed jet geometries.

\begin{acknowledgements}
    This publication makes use of data products from the TRAPPIST project. TRAPPIST-South is funded by the Belgian National Fund for Scientific Research (F.R.S.-FNRS) under grant PDR T.0120.21. TRAPPIST-North is funded by the University of Li\`{e}ge, and performed in collaboration with Cadi Ayyad University of Marrakesh. E. Jehin is a FNRS Senior Research Associates and J. Manfroid is a Honorary Research Director of the FNRS. This publication benefits from the support of the French Community of Belgium in the context of the FRIA Doctoral Grant awarded to L. A. Head. 
\end{acknowledgements}

\bibliography{main}

\begin{appendix}
    \include{supplementary_material}

\end{appendix}

\end{document}

%% file: supplementary_material.tex
\renewcommand{\thefigure}{S\arabic{figure}}
\renewcommand{\thetable}{S\arabic{table}}
\onecolumn

\section{Supplementary tables}

\begin{table*}[h!]
    \begin{center}
         \caption{
            Table of all observations made in 2014-2015. 
        }
        \label{tab:obs_2014}
        \smallskip
        \begin{threeparttable}
       
        
            \begin{tabular}{| c | c | c | c | c | c | c | c |}
                \hline
                & Date & Detected & 	Telescope & Observation time &		 $\phi_S$ (\textdegree) &	$\phi_E$ (\textdegree) & $\theta_{S3}$ (\textdegree) \\
                    
                \hline
    
                1     & 2014-12-04 & Yes   & TS    & 08:17:20 - 08:23:53 & -93   & -102  & 143 \\
                2     & 2014-12-06 & Yes   & TS    & 06:25:29 - 08:34:44 & -53   & -62   & 271 \\
                3     & 2014-12-08 & No    & TS    & 06:05:19 - 06:35:13 & -16   & -25   & 49 \\
                4     & 2014-12-12 & No    & TS    & 07:41:27 - 08:25:15 & 92    & 83    & 212 \\
                5     & 2014-12-19 & No    & TS    & 06:51:07 - 07:05:15 & 66    & 58    & 253 \\
                6     & 2014-12-21 & Yes   & TS    & 08:11:39 - 08:30:36 & 125   & 117   & 321 \\
                7     & 2015-01-11 & Yes   & TS    & 07:40:34 - 08:00:03 & 72    & 68    & 5 \\
                8     & 2015-01-13 & No    & TS    & 05:57:56 - 07:25:27 & 109   & 106   & 142 \\
                9     & 2015-01-20 & No    & TS    & 07:43:08 - 08:00:41 & 103   & 100   & 121 \\
                10    & 2015-01-23 & No    & TS    & 07:46:51 - 08:09:28 & -6    & -8    & 278 \\
                11    & 2015-01-28 & Yes   & TS    & 07:22:02 - 07:45:40 & -72   & -72   & 194 \\
                12    & 2015-02-01 & No    & TS    & 05:55:08 - 06:30:16 & 10    & 10    & 83 \\
                13    & 2015-02-12 & Yes   & TS    & 02:00:10 - 02:19:50 & 53    & 55    & 57 \\
                14    & 2015-03-29 & Yes   & TS    & 03:49:33 - 04:10:26 & -137  & -128  & 232 \\
                15    & 2015-03-30 & Yes   & TS    & 03:36:23 - 04:10:04 & 65    & 74    & 287 \\
                16    & 2015-03-31 & Yes   & TS    & 03:55:21 - 04:10:16 & -90   & -80   & 336 \\
                17    & 2015-04-01 & Yes   & TS    & 00:02:22 - 00:30:21 & 81    & 91    & 132 \\
    
                \hline
            \end{tabular}
            \begin{tablenotes}
                \footnotesize \item The ``Detected'' column indicates whether the presence of a jet-like structure was identified on this date. \textbf{``TS'' refers to the telescope used for the observation, TRAPPIST-South (IAU code I40).} \textbf{The ``Observation time'' column gives the observation period in UTC.} $\phi_S$ is the average orbital angle of Io in images taken on this date, with 0\textdegree\ placing Io behind Jupiter with respect to the Sun, \textbf{with positive orbital angles moving Io in an anti-clockwise direction}. $\phi_E$ is the average Earth-Jupiter-Io angle. $\theta_{S3}$ is the average magnetic longitude of Io in System-III.
            \end{tablenotes}
        
        \end{threeparttable}
    \end{center}
    
\end{table*}

\begin{table*}[h!] 
    
    \begin{center}
        \caption{
            Table of all observations made in 2021.
        }
        \label{tab:obs_2021}
        \smallskip
        \begin{threeparttable}
            \begin{tabular}{| c | c | c | c | c | c | c | c | }
                \hline
    
                & Date & Detected & 	Telescope & Observation time &		 $\phi_S$ (\textdegree) &	$\phi_E$ (\textdegree) & $\theta_{S3}$ (\textdegree) \\
     
                \hline

                1     & 2021-04-28 & No    & TS    & 09:31:03 - 09:42:34 & 87    & 76    & 20 \\
                2     & 2021-04-29 & No    & TS    & 10:01:57 - 10:09:48 & -66   & -77   & 59 \\
                3     & 2021-05-14 & Yes   & TS    & 10:10:55 - 10:20:09 & 107   & 96    & 128 \\
                4     & 2021-05-15 & No    & TS    & 10:08:02 - 10:17:16 & -50   & -62   & 182 \\
                5     & 2021-05-22 & No    & TS    & 10:16:22 - 10:26:04 & -65   & -77   & 188 \\
                6     & 2021-05-29 & No    & TS    & 10:17:49 - 10:32:40 & -81   & -92   & 198 \\
                7     & 2021-05-30 & No    & TS    & 10:16:56 - 10:30:28 & 123   & 112   & 252 \\
                8     & 2021-05-31 & No    & TS    & 10:16:50 - 10:30:11 & -35   & -46   & 303 \\
                9     & 2021-06-05 & No    & TS    & 10:17:47 - 10:28:14 & -97   & -108  & 208 \\
                10    & 2021-06-06 & Yes   & TS    & 10:14:17 - 10:30:01 & 107   & 96    & 261 \\
                11    & 2021-06-07 & Yes   & TS    & 10:06:06 - 10:15:09 & -53   & -63   & 319 \\
                12    & 2021-06-12 & No    & TS    & 10:21:27 - 10:33:36 & -113  & -123  & 216 \\
                13    & 2021-06-13 & No    & TS    & 10:24:19 - 10:34:46 & 91    & 81    & 268 \\
                14    & 2021-06-20 & No    & TS    & 10:31:20 - 10:39:59 & 76    & 66    & 276 \\
                15    & 2021-06-23 & No    & TN    & 04:05:47 - 04:11:27 & -89   & -98   & 253 \\
                16    & 2021-06-24 & Yes    & TN    & 04:03:32 - 04:07:49 & 115   & 105   & 308 \\
                17    & 2021-06-28 & Yes   & TS    & 10:11:22 - 10:24:24 & -100  & -109  & 347 \\
                18    & 2021-06-30 & No    & TS    & 10:08:01 - 10:17:29 & -103  & -111  & 255 \\
                19    & 2021-06-30 & No    & TN    & 04:23:36 - 04:28:06 & -54   & -63   & 95 \\
                20    & 2021-07-01 & No    & TN    & 04:23:24 - 04:30:15 & 101   & 93    & 309 \\
                21    & 2021-07-02 & Yes   & TN    & 04:33:27 - 04:42:43 & -55   & -63   & 356 \\
                22    & 2021-07-04 & No    & TS    & 10:16:12 - 10:25:24 & 41    & 33    & 303 \\
                23    & 2021-07-08 & No    & TS    & 09:02:04 - 09:12:09 & 125   & 117   & 190 \\
                24    & 2021-07-15 & No    & TS    & 10:11:26 - 10:20:57 & 118   & 112   & 168 \\
                25    & 2021-07-16 & No    & TN    & 04:02:13 - 04:23:28 & -91   & -97   & 29 \\
                26    & 2021-07-23 & Yes   & TS    & 10:01:25 - 10:11:18 & -57   & -61   & 235 \\
                27    & 2021-09-26 & No    & TS    & 23:32:13 - 23:42:05 & -41   & -32   & 62 \\
                28    & 2021-10-01 & No    & TS    & 23:46:43 - 23:56:42 & -102
      & -93 & 320 \\
                29    & 2021-10-02 & No    & TS    & 23:36:32 - 23:42:48 & 100 &  	110 & 19 \\
                30    & 2021-10-03 & No    & TS    & 23:36:30 - 23:45:28 & -57&  	-47 & 71 \\
                31   & 2021-10-10 & No    & TS    & 00:00:11 - 00:19:47 &  89 & 	99 & 16 \\

                \hline

            \end{tabular}
            \begin{tablenotes}
                \footnotesize \item Column titles are identical to those of Table \ref{tab:obs_2014}. ``TN'' refers to the TRAPPIST-North telescope (IAU code Z53).
            \end{tablenotes}
        \end{threeparttable}
    \end{center}
    
\end{table*}

\begin{table*}[h!]
    \begin{center}
        \caption{
            Table of all observations made in 2022-2023.
        }
        \label{tab:obs_2022}
        \smallskip
        \begin{threeparttable}
            
            \begin{tabular}{| c | c | c | c | c | c | c | c | }
                \hline
                & Date & Detected & 	Telescope & Observation time &		 $\phi_S$ (\textdegree) &	$\phi_E$ (\textdegree) & $\theta_{S3}$ (\textdegree) \\

                \hline

                1     & 2022-05-26 & Yes   & TS    & 09:13:37 - 09:52:37 & 102   & 92    & 312 \\
                2     & 2022-06-19 & No    & TS    & 09:43:36 - 10:10:06 & -52   & -64   & 130 \\
                3     & 2022-07-03 & Yes   & TS    & 10:02:30 - 10:37:21 & -82   & -94   & 141 \\
                4     & 2022-08-19 & Yes   & TS    & 10:02:01 - 10:19:27 & 117   & 110   & 111 \\
                5     & 2022-08-20 & No    & TS    & 08:27:26 - 08:38:49 & -54   & -60   & 207 \\
                6     & 2022-09-03 & Yes   & TS    & 09:21:35 - 09:44:56 & -78   & -82   & 201 \\
                7     & 2022-09-04 & No    & TS    & 05:02:55 - 09:20:00 & 124   & 120   & 263 \\
                8     & 2022-09-05 & Yes   & TS    & 04:47:51 - 05:14:55 & -71   & -74   & 77 \\
                9     & 2022-09-14 & No    & TN    & 02:20:08 - 02:30:52 & -61   & -63   & 262 \\
                10    & 2022-09-27 & No    & TS    & 02:01:57 - 02:20:55 & 62    & 63    & 238 \\
                11    & 2022-11-23 & Yes   & TN    & 18:32:20 - 18:45:06 & -86   & -75   & 280 \\
                12    & 2022-11-27 & Yes   & TS    & 02:16:25 - 02:35:33 & -130  & -119  & 222 \\
                13    & 2022-11-27 & No    & TN    & 23:33:10 - 23:45:12 & 51    & 62    & 353 \\
                14    & 2022-11-29 & No    & TN    & 20:11:30 - 20:23:30 & 70    & 81    & 193 \\
                15    & 2022-12-21 & No    & TN    & 21:03:02 - 21:14:29 & -129  & -118  & 254 \\
                16    & 2022-12-22 & No    & TN    & 18:47:04 - 18:59:19 & 56    & 67    & 11 \\
                17    & 2022-12-23 & No    & TN    & 18:47:11 - 18:59:26 & -102  & -90   & 63 \\
                18    & 2022-12-24 & No    & TN    & 19:02:15 - 19:14:19 & 104   & 116   & 110 \\
                19    & 2023-01-06 & No    & TS    & 00:56:58 - 01:19:57 & 76    & 87    & 219 \\
                20    & 2023-01-07 & No    & TN    & 18:52:27 - 19:04:25 & 71    & 82    & 137 \\
                21    & 2023-01-14 & No    & TN    & 19:05:44 - 19:17:37 & 56    & 67    & 142 \\
                22    & 2023-01-15 & No    & TN    & 19:01:52 - 19:14:06 & -101  & -91   & 196 \\
                23    & 2023-01-16 & No    & TN    & 18:52:15 - 19:04:39 & 101   & 111   & 254 \\

                \hline

            \end{tabular}
            \begin{tablenotes}
                \footnotesize \item Column titles are identical to those of Tables \ref{tab:obs_2014} and \ref{tab:obs_2021}.
            \end{tablenotes}
        \end{threeparttable}
    \end{center}
    
\end{table*}

\FloatBarrier
\section{Supplementary figures}
\begin{figure*}[h!]
    \begin{center}
        \includegraphics[height=0.92\textheight,keepaspectratio]{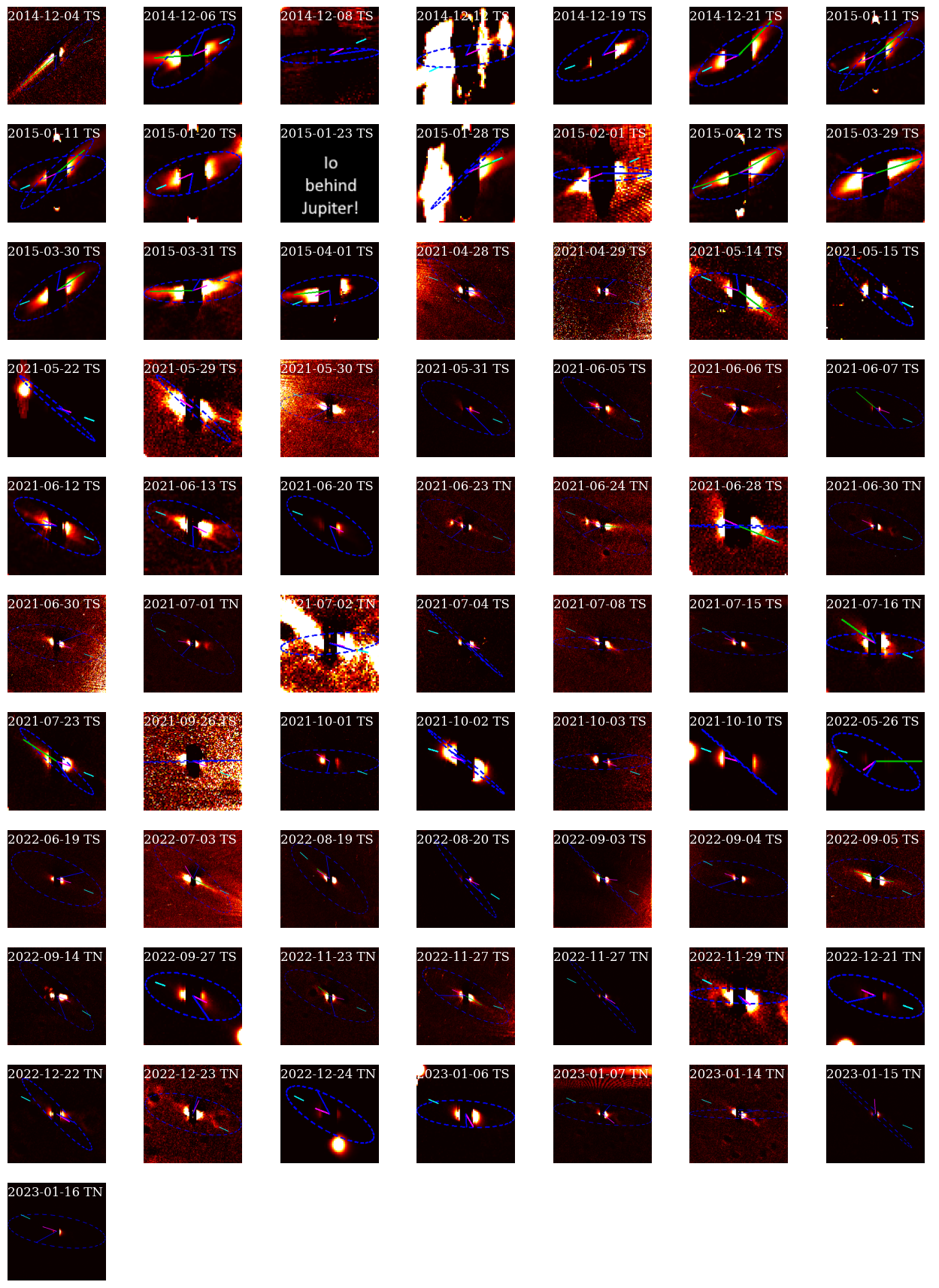}
    \end{center}
\end{figure*}
\setcounter{figure}{0}   
\begin{figure*}[h!]
    \caption{
        Stacked images of all TRAPPIST viewing intervals presented in this work, cropped around Io and having undergone the processing described in Sect. 2.2 of the main article. Images are orientated with north upward and west to the right. The direction toward the centre of Jupiter in the image is indicated by a short cyan line. The short magenta line centred at Io indicates the apparent direction of movement of Io in the image. The dashed blue ellipse represents the plane perpendicular to the local magnetic field at Io according to an observer on Earth, and the blue line between the centre of Io and the edge of this ellipse is the projection of the movement vector of Io in this plane. Green lines represent detected jet-like structures. On 2015-01-23, Io was eclipsed by Jupiter and hence not visible.
    }
    \label{fig_SM:mosaic}
\end{figure*}

\begin{figure*}[h!]
    \begin{center}
        \subfloat[2014-12-06 ($\theta_{S3}$ = 271\textdegree)]{
            \includegraphics[width=\textwidth]{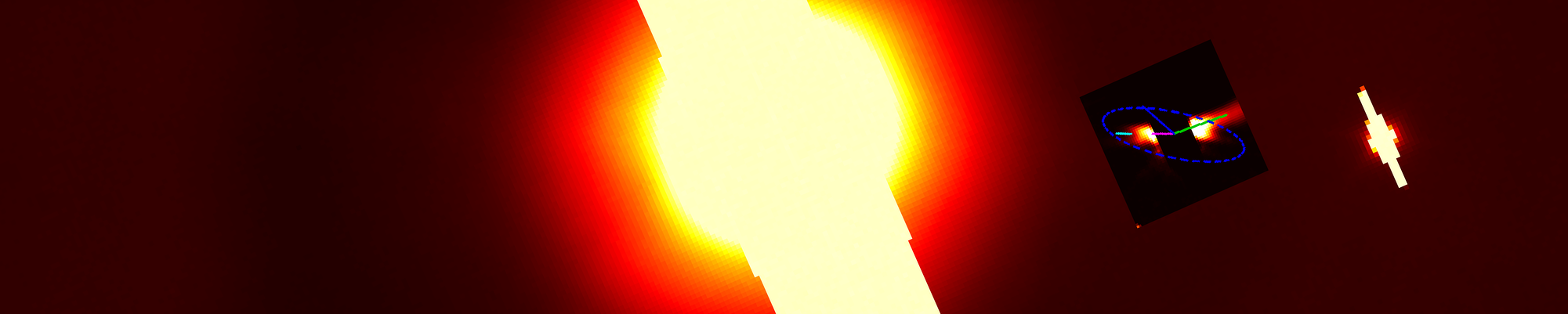}
        }\\
        \subfloat[2021-05-31 ($\theta_{S3}$ = 303\textdegree)]{
            \includegraphics[width=\textwidth]{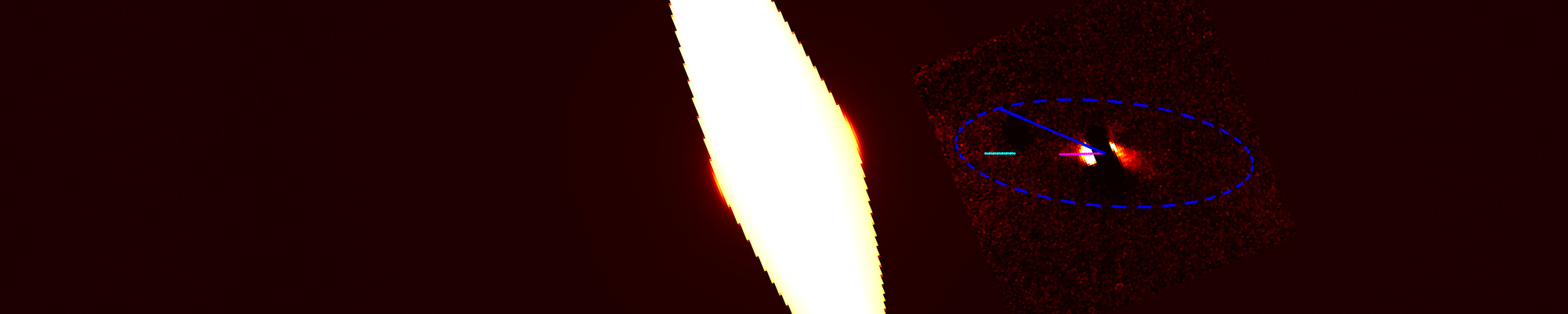}
        }
    \end{center}
    \caption{Comparison of Io in two cases with similar System-III longitudes. Images are orientated with north upward and west to the right. The direction toward the centre of Jupiter in the image is indicated by a short cyan line. The short magenta line centred at Io indicates the apparent direction of movement of Io in the image. The dashed blue ellipse represents the plane perpendicular to the local magnetic field at Io according to an observer on Earth, and the blue line between the centre of Io and the edge of this ellipse is the projection of the movement vector of Io in this plane. Green lines represent detected jet-like structures.}
    \label{fig_SM:S3}
\end{figure*}

\begin{figure*}[h!]
    \begin{center}
        \subfloat[2014-12-21 ($\phi_{E}$ = 117\textdegree)]{
            \includegraphics[width=\textwidth]{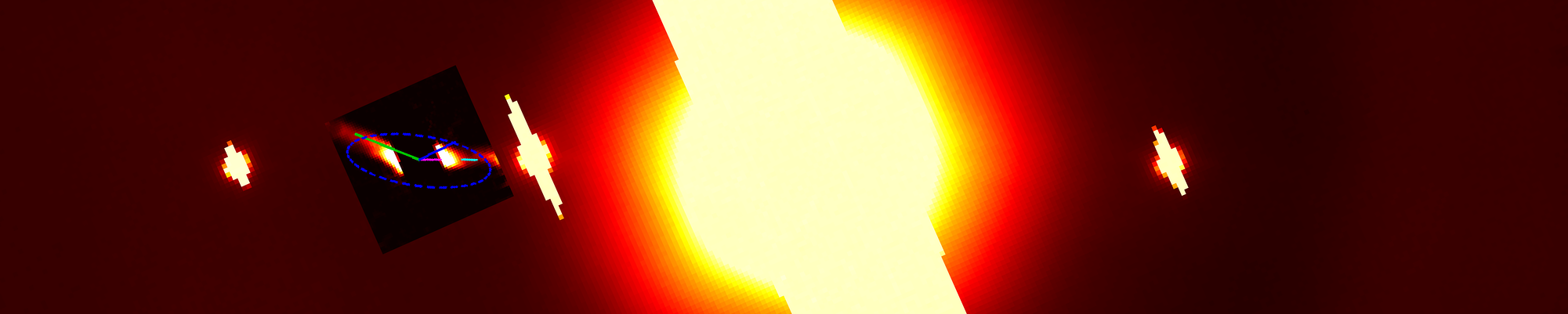}
        }\\
        \subfloat[2021-07-08 ($\phi_{E}$ = 117\textdegree)]{
            \includegraphics[width=\textwidth]{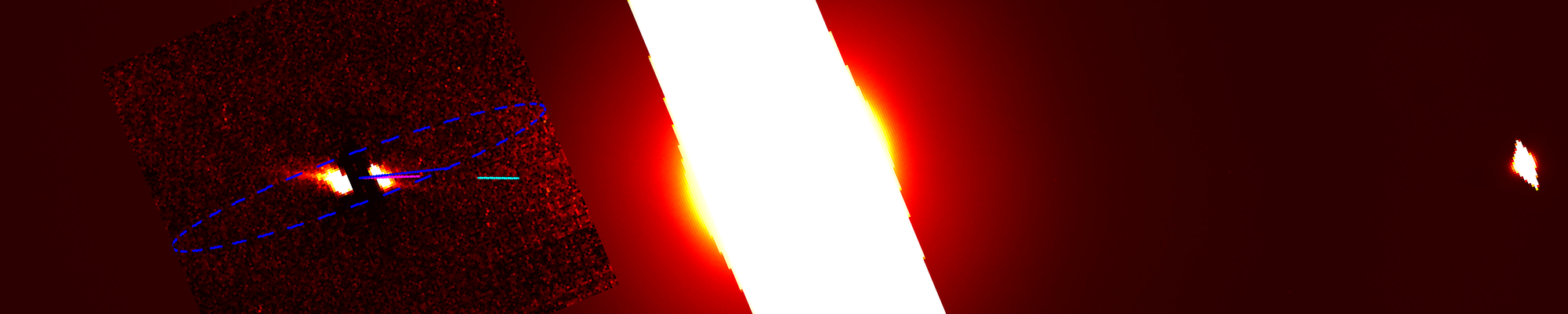}
        }
    \end{center}
    \caption{Comparison of Io in two cases with similar Earth-Jupiter-Io phase angles. Images are orientated with north upward and west to the right. The direction toward the centre of Jupiter in the image is indicated by a short cyan line. The short magenta line centred at Io indicates the apparent direction of movement of Io in the image. The dashed blue ellipse represents the plane perpendicular to the local magnetic field at Io according to an observer on Earth, and the blue line between the centre of Io and the edge of this ellipse is the projection of the movement vector of Io in this plane. Green lines represent detected jet-like structures.}
    \label{fig_SM:PA}
\end{figure*}

\begin{figure*}[h!]
    \begin{center}
        \subfloat[2021-06-23]{
            \includegraphics[width=\textwidth]{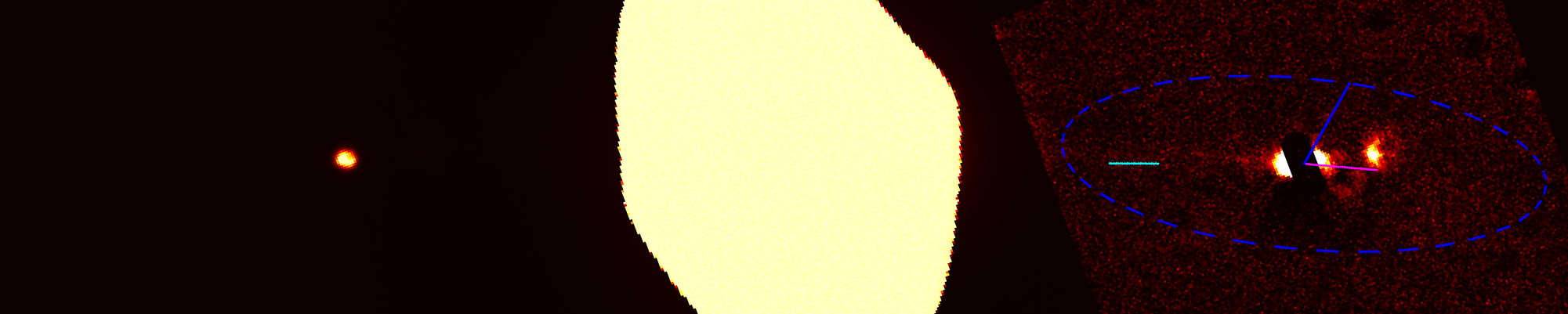}
        }\\
        \subfloat[2021-06-24]{
            \includegraphics[width=\textwidth]{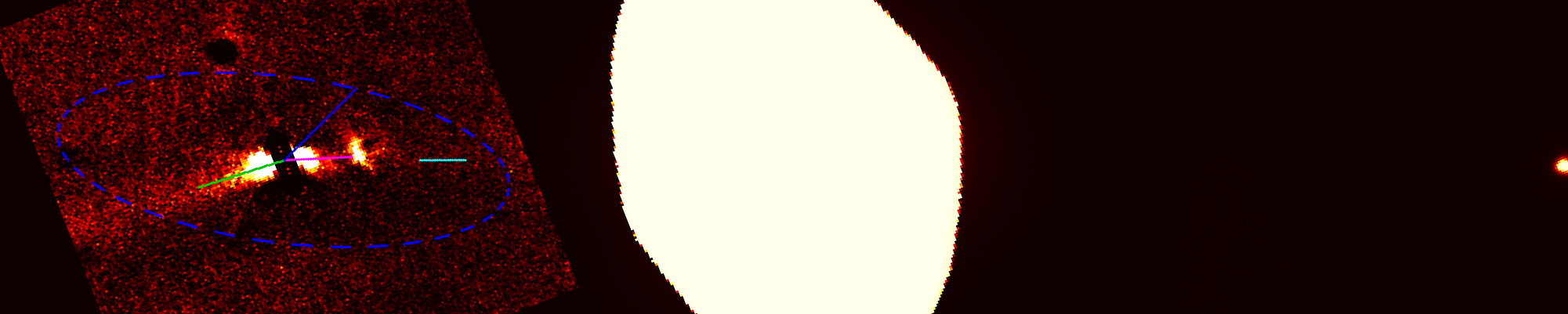}
        }
    \end{center}
    \caption{Comparison of Io over an interval of one day. Images are orientated with north upward and west to the right. The direction toward the centre of Jupiter in the image is indicated by a short cyan line. The short magenta line centred at Io indicates the apparent direction of movement of Io in the image. The dashed blue ellipse represents the plane perpendicular to the local magnetic field at Io according to an observer on Earth, and the blue line between the centre of Io and the edge of this ellipse is the projection of the movement vector of Io in this plane. Green lines represent detected jet-like structures.}
    \label{fig_SM:1day_diff}
\end{figure*}

\begin{figure*}[h!]
    \begin{center}
        \subfloat[2015-01-11 ($\theta_{S3}$ = 5\textdegree, $\phi_{E}$ = 68\textdegree)]{
            \includegraphics[width=\textwidth]{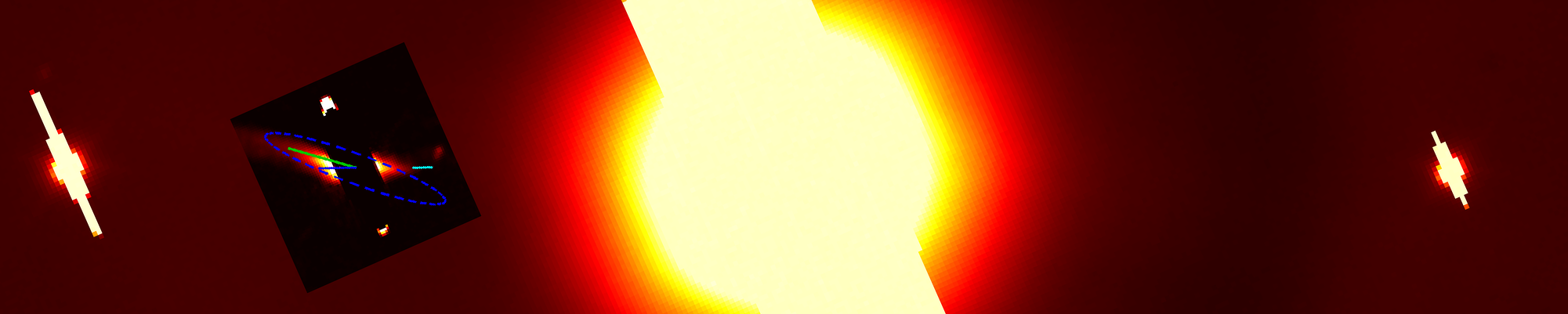}
        }\\
        \subfloat[2022-12-22 ($\theta_{S3}$ = 11\textdegree, $\phi_{E}$ = 67\textdegree)]{
            \includegraphics[width=\textwidth]{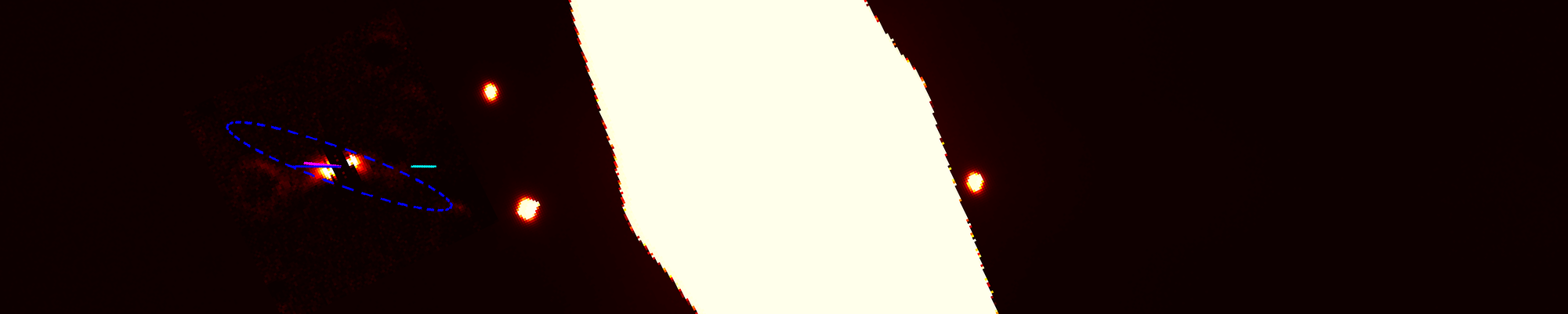}
        }
    \end{center}
    \caption{Comparison of Io in two cases with similar System-III longitudes and Earth-Jupiter-Io phase angles. Images are orientated with north upward and west to the right. The direction toward the centre of Jupiter in the image is indicated by a short cyan line. The short magenta line centred at Io indicates the apparent direction of movement of Io in the image. The dashed blue ellipse represents the plane perpendicular to the local magnetic field at Io according to an observer on Earth, and the blue line between the centre of Io and the edge of this ellipse is the projection of the movement vector of Io in this plane. Green lines represent detected jet-like structures.}
    \label{fig_SM:S3_and_PA}
\end{figure*}

%% file: main.bbl
\begin{thebibliography}{27}
\expandafter\ifx\csname natexlab\endcsname\relax\def\natexlab#1{#1}\fi

\bibitem[{Bagenal \& Dols(2020)}]{bagenalanddols2020}
Bagenal, F. \& Dols, V. 2020, Journal of Geophysical Research: Space Physics,
  125, e2019JA027485

\bibitem[{Bergstralh {et~al.}(1975)Bergstralh, Matson, \&
  Johnson}]{bergstralhEA1975}
Bergstralh, J.~T., Matson, D.~L., \& Johnson, T.~V. 1975, Astrophysical
  Journal, 195, 131

\bibitem[{Burger \& Johnson(2004)}]{burgerandjohnson2004}
Burger, M.~H. \& Johnson, R.~E. 2004, Icarus, 171, 557

\bibitem[{Connerney {et~al.}(2020)Connerney, Timmins, Herceg, \&
  Joergensen}]{connerneyEA2020}
Connerney, J. E.~P., Timmins, S., Herceg, M., \& Joergensen, J.~L. 2020,
  Journal of Geophysical Research: Space Physics, 125, e2020JA028138

\bibitem[{Connerney {et~al.}(2022)}]{connerneyEA2022}
Connerney, J. E.~P. {et~al.} 2022, Journal of Geophysical Research: Planets,
  127, e2021JE007055

\bibitem[{de~Kleer {et~al.}(2019)de~Kleer, Nimmo, \& Kite}]{dekleerEA2020}
de~Kleer, K., Nimmo, F., \& Kite, E. 2019, Geophysical Research Letters, 46,
  6327

\bibitem[{de~Pater {et~al.}(2020)de~Pater, Luszcz-Cook, Rojo, Redwing,
  de~Kleer, \& Moullet}]{depaterEA2020}
de~Pater, I., Luszcz-Cook, S., Rojo, P., {et~al.} 2020, Planetary Science
  Journal, 1

\bibitem[{de~Spiegeleire(2019)}]{spiegeleire2019}
de~Spiegeleire, A. 2019, Master's thesis, University of Li\`{e}ge, Li\`{e}ge

\bibitem[{Dols {et~al.}(2008)Dols, Delamere, \& Bagenal}]{dolsEA2008}
Dols, V., Delamere, P.~A., \& Bagenal, F. 2008, Journal of Geophysical
  Research: Space Physics, 113, A09208

\bibitem[{Geissler \& Goldstein(2007)}]{geisslerandgoldstein2007}
Geissler, P.~E. \& Goldstein, D.~B. 2007, in Io after Galileo, ed. R.~M.~C.
  Lopez \& J.~R. Spencer (Heidelberg: Springer Berlin)

\bibitem[{Grava {et~al.}(2021)}]{gravaEA2021}
Grava, C. {et~al.} 2021, The Astronomical Journal, 162, 190

\bibitem[{James {et~al.}(2022)James, Provan, Kamran, Wilson, Vogt, Brennan, \&
  Cowley}]{jupitermag}
James, M.~K., Provan, G., Kamran, A., {et~al.} 2022, Jupiter{M}ag,
  \url{https://github.com/mattkjames7/JupiterMag.git}, version 1.0.8

\bibitem[{Jehin {et~al.}(2011)Jehin, Gillon, Queloz, Magain, Manfroid, Chantry,
  Lendl, Hutsemekers, \& Udry}]{jehinEA2011}
Jehin, E., Gillon, M., Queloz, D., {et~al.} 2011, The Messenger, 145

\bibitem[{Lellouch(2005)}]{lellouch2005}
Lellouch, E. 2005, Space Science Reviews, 116, 211

\bibitem[{Lellouch {et~al.}(2007)Lellouch, McGrath, \& Jessup}]{lellouchEA2007}
Lellouch, E., McGrath, M.~A., \& Jessup, K.~L. 2007, in Io after Galileo, ed.
  R.~M.~C. Lopez \& J.~R. Spencer (Heidelberg: Springer Berlin)

\bibitem[{Mendillo {et~al.}(1990)Mendillo, Baumgardner, Flynn, \&
  Hughes}]{mendilloEA1990}
Mendillo, M., Baumgardner, J., Flynn, B., \& Hughes, W.~J. 1990, Nature, 348,
  312

\bibitem[{Roth {et~al.}(2020)}]{rothEA2020}
Roth, L. {et~al.} 2020, Icarus, 350, 113925

\bibitem[{Schneider \& Bagenal(2007)}]{schneiderandbagenal2007}
Schneider, N. \& Bagenal, F. 2007

\bibitem[{Schneider {et~al.}(1991)Schneider, Trauger, Wilson, Brown, Evans, \&
  Shemansky}]{schneiderEA1991}
Schneider, N.~M., Trauger, J.~T., Wilson, J.~K., {et~al.} 1991, Science, 253,
  1394

\bibitem[{Smyth(1992)}]{smyth1992}
Smyth, W.~H. 1992, Advances in Space Research, 12, 337

\bibitem[{Summers {et~al.}(1989)Summers, Strobel, Yung, Trauger, \&
  Mills}]{summersEA1989}
Summers, M.~E., Strobel, D.~F., Yung, Y.~L., Trauger, J.~T., \& Mills, F. 1989,
  Astrophysical Journal, 343, 468

\bibitem[{Thomas {et~al.}(2004)Thomas, Bagenal, Hill, \& Wilson}]{thomasEA2004}
Thomas, N., Bagenal, F., Hill, T.~W., \& Wilson, J.~K. 2004, in Jupiter: The
  Planets, Satellites and Magnetosphere, ed. F.~Bagenal, T.~Dowling, \&
  W.~McKinnon (Cambridge: Cambridge University Press)

\bibitem[{Williams \& Howell(2007)}]{williamsandhowell2007}
Williams, D.~A. \& Howell, R.~R. 2007, in Io after Galileo, ed. R.~M.~C. Lopez
  \& J.~R. Spencer (Heidelberg: Springer Berlin)

\bibitem[{Wilson {et~al.}(2002)Wilson, Mendillo, Baumgardner, Schneider,
  Trauger, \& Flynn}]{wilsonEA2002}
Wilson, J.~K., Mendillo, M., Baumgardner, J., {et~al.} 2002, Icarus, 157, 476

\bibitem[{Yoneda {et~al.}(2015)Yoneda, Kagitani, Tsuchiya, Sakanoi, \&
  Okano}]{yonedaEA2015}
Yoneda, M., Kagitani, M., Tsuchiya, F., Sakanoi, T., \& Okano, S. 2015, Icarus,
  261, 31

\bibitem[{Yoshikawa {et~al.}(2017)}]{yoshikawaEA2017}
Yoshikawa, I. {et~al.} 2017, Earth, Planets and Space, 69

\bibitem[{Yoshioka {et~al.}(2018)}]{yoshiokaEA2018}
Yoshioka, K. {et~al.} 2018, Geophysical Research Letters, 45, 10193

\end{thebibliography}
